\documentclass[%
reprint,
superscriptaddress,
nofootinbib,
amsmath,amssymb,
aps,
pre,
notitlepage,
showkeys]{revtex4-2}
\usepackage[colorlinks=true, urlcolor=blue, anchorcolor=blue, citecolor=blue,filecolor=blue,linkcolor=blue,menucolor=blue]{hyperref}
\usepackage{url}
\usepackage{graphicx}
\usepackage{lipsum}
\usepackage[export]{adjustbox}
\usepackage{mathtools}
\usepackage[T1]{fontenc}
\usepackage{amsmath}
\usepackage{amssymb}
\usepackage{stmaryrd}
\usepackage[utf8]{inputenc}
\usepackage{rotating}
\usepackage{comment}
\usepackage{url}
\usepackage{bbm}
\usepackage{soul}
\usepackage[normalem]{ulem}
\usepackage{tabularx,booktabs}
\usepackage{stmaryrd}
\newcolumntype{Y}{>{\centering\arraybackslash}X}

\usepackage{hyperref}  
\usepackage{subfigure}
\usepackage{xr}
\usepackage{tcolorbox}
\usepackage{paralist}
\usepackage{booktabs}
\usepackage{xcolor}
\usepackage{pifont}

\definecolor{myblue}{RGB}{119, 170, 218}
\definecolor{mygreen}{RGB}{44, 126, 51}

\newcommand{\cmark}{\textcolor{green!60!black}{\ding{51}}}
\newcommand{\xmark}{\textcolor{red!70!black}{\ding{55}}}

\begin{document}
\title{A generative model of function growth explains hidden self-similarities across biological and social systems} 
\author{James Holehouse}
\email{jamesholehouse1@gmail.com}
\affiliation{The Santa Fe Institute, 1399 Hyde Park Road, Santa Fe, NM, 87510, USA.}

\author{S.~Redner}
\affiliation{The Santa Fe Institute, 1399 Hyde Park Road, Santa Fe, NM, 87510, USA.}

\author{Vicky Chuqiao Yang}
\affiliation{Sloan School of Management, Massachusetts Institute of Technology, Cambridge MA, USA.}

\author{P.~L.~Krapivsky}
\affiliation{The Santa Fe Institute, 1399 Hyde Park Road, Santa Fe, NM, 87510, USA.}
\affiliation{Department of Physics, Boston University, Boston, MA 02215, USA.}

\author{Jos\'e Ignacio Arroyo}
\affiliation{The Santa Fe Institute, 1399 Hyde Park Road, Santa Fe, NM, 87510, USA.}

\author{Geoffrey B.~West}
\affiliation{The Santa Fe Institute, 1399 Hyde Park Road, Santa Fe, NM, 87510, USA.}

\author{Chris Kempes}
\affiliation{The Santa Fe Institute, 1399 Hyde Park Road, Santa Fe, NM, 87510, USA.}

\author{Hyejin Youn}
\affiliation{The Santa Fe Institute, 1399 Hyde Park Road, Santa Fe, NM, 87510, USA.}
\affiliation{Graduate School of Business, Seoul National University, Seoul, South Korea.}
\affiliation{Northwestern Institute on Complex Systems, Evanston, IL, USA.}

\begin{abstract}
    From genomes and ecosystems to bureaucracies and cities, the growth of complex systems occurs by adding new types of functions and expanding existing ones. We present a simple generative model that generalizes the Yule–Simon process by including: (i) a size-dependent probability of introducing new functions, and (ii) a generalized preferential attachment mechanism for expanding existing ones. We uncover a shared underlying structure that helps explain how function diversity evolves in empirical observations, such as prokaryotic proteomes, U.S.~federal agencies, and urban economies. We show that real systems are often best represented as having non-Zipfian rank-frequency distributions, driven by sublinear preferential attachment, whilst still maintaining power-law scaling in their abundance distributions. Furthermore, our analytics explain five distinct phases of the organization of functional elements across complex systems. The model integrates empirical findings regarding the logarithmic growth of diversity in cities and the self-similarity of their rank-frequency distributions. Self-similarity previously observed in the rank-frequency distributions of cities is not observed in cells and federal agencies---however, under a rescaling relative to the total diversity, all systems admit self-similar structures predicted by our theory.
    
\end{abstract}

\maketitle



\section*{Introduction}

\noindent Diversity is an essential characteristic of a complex system, often defined by the number of elements with distinct functions that comprise the system. The important consequences of diversity have recently been studied in a variety of biological, social, and ecological contexts \cite{mayr1997challenge,page2010diversity,bouchaud2013crises,chase2011disentangling,lamanna2017plant,scheffer2015generic,tikhonov2017collective,staps2019emergence,marquez2025nature,youn2016scaling,piantadosi2014zipf}. Diversity in ecosystems gives rise to resilience \cite{ives2007stability}, diversity in teams giving rise to increased team performance \cite{page2008difference}, and a diversity of chemical species protects auto-catalytic chemical reaction networks from crashes \cite{mehrotra2009diversity}. The type of diversity exhibited in these systems is functional in nature and is distinct from diversity arising from system noise, e.g., sample noise and compositional variability \cite{morrison2025quantifying} or intrinsic and extrinsic stochastic fluctuations \cite{elowitz2002stochastic}. The increase in diversity with the expansion of a system has been suggested to constitute a biological law, known as the \textit{zero-force evolutionary law} \cite{mcshea2010biology}, and similar claims have been made in social systems \cite{youn2016scaling,mazzarisi2021maximal}. The perceived importance of diversity for the success and resilience of complex systems strongly suggests the need to develop better models for the growth of diversity in a system. In this paper, we address this issue by introducing a simple, yet general, model that captures many of the essential underlying features of the dynamics of diversity.

General mechanisms have been proposed for explaining a variety of shapes in diversity distributions, ranging from the random sampling of the Yule-Simon model, to the power-law scaling of diversity implied by Heaps' law, to preferential attachment \cite{yule1925ii,simon1955class,barabasi1999emergence,van2005formal,hebert2012structural,loreto2016dynamics,taalbi2025long}. More specific mechanisms have also been proposed, such as processes of evolutionary innovation where newly evolved elements are retained or discarded based on fitness \cite{hochberg2017innovation,montevil2019possibility}. These more specific mechanisms rely on the idea that diverse structures evolve to be functionally optimal \cite{west1997general} and robust in a variety of expected environments \cite{krakauer2005principles,wagner2008robustness}. Given the increase in the number of approaches for modeling the growth of diversity, the simplest and most useful narratives still focus on the diversity of elements and the relative abundance among distinct elements---which is what makes the seminal approaches of Yule and Simon still so popular \cite{yule1925ii,yule2014statistical,simon1955class}.

\begin{figure}[ht]
    \includegraphics[width=.5\textwidth]{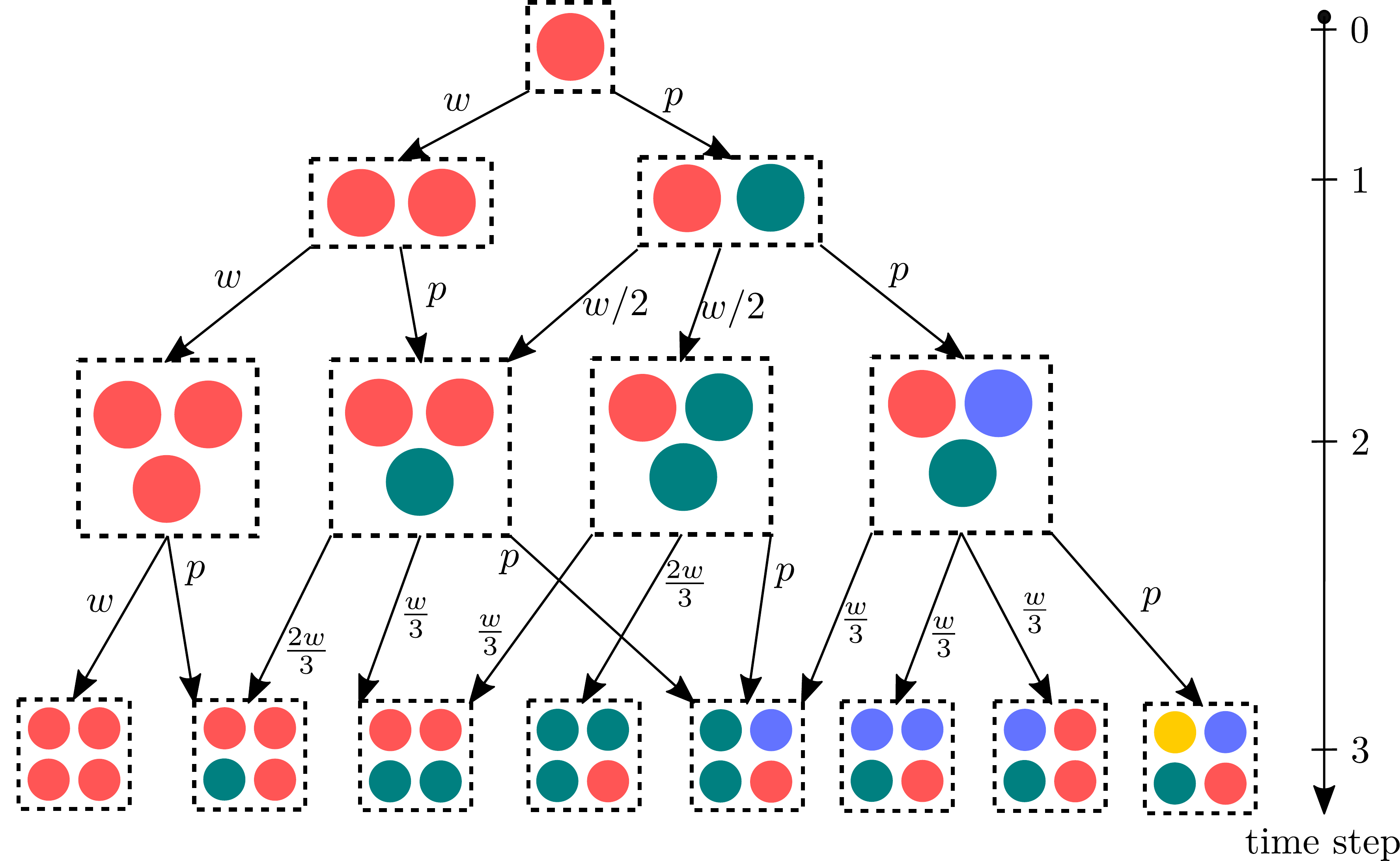}
    \caption{\textbf{The first three time steps for the Yule-Simon model.} Here, we denote $w=1-p$. Each function subscribes to linear preferential attachment and new functions created at a constant rate. Different colors represent different functions and a single worker is added at each update.}
    \label{fig:standardYS}
\end{figure}

Simon's model makes two key assumptions. First, it assumes that the rate of expansion of any existing function is proportional to its size. Second, it assumes that there is a constant probability of adding a new functional type to the system (see Fig.~\ref{fig:standardYS}). This results in the dynamical equation
\begin{align}\label{eq:genME}
    \frac{\partial n_k}{\partial N}  = p \delta_{k,1} + (1-p)\left(q_{k-1} n_{k-1}-q_k n_{k}\right),
\end{align}
where $n_k = n_k(N)$ is the number of functions of size $k$ for system size $N$. At each time step the system size $N$ increases by 1, $p$ is the probability to add a new function, and $q_k = k/N$. We have assumed that $N$ is large enough such that $n_k(N+1)-n_k(N)\approx \partial n_k(N)/\partial N$. Simon's model admits an analytical solution that is extensive, i.e., scales linearly with $N$, and has an algebraic large $k$ tail 
$$n_k\sim N k^{-(1+1/(1-p))}$$
which captures the Zipfian behavior of power-law distributed system components in the limit of large $k$. 
However, it fails to capture power-law dynamics of diversity growth seen in a variety of systems \cite{mazzolini2018zipf}. 
The number of unique elements versus system size, $D(N)$, often scales as a power law in many systems, an empirical result known as Heaps' law; however, in some systems, such as cities \cite{youn2016scaling} or ecosystems \cite{gleason1922relation}, it scales logarithmically with $N$. 

Thus, while diversity patterns are useful for our general understanding of complex systems, especially as connected with simple generative models, it has also been widely recognized that existing models fail to capture a variety of data from different domains suggesting that the organization of many networks is not Zipfian \cite{stumpf2012critical,loreto2016dynamics,broido2019scale,artico2020rare,holme2019rare}. For example, in addition to the logarithmic growth of diversity in cities \cite{youn2016scaling} and ecosystems \cite{gleason1922relation} 
the origins of the self-similar organization of cities---not observed across other complex systems---remains an open problem \cite{youn2016scaling,yang2025principles}.

Here we develop a more general model that accounts for the two critical aspects of generating diversity from a fixed set of elements: (i) that the propensity for a system to add new functions decreases with increasing system size \cite{van2005formal,okie2025equilibrium,yang2025principles}; and (ii) that larger functions grow faster, but not necessarily in proportion to their size, as has been observed across a variety of systems \cite{newman2001clustering,peltomaki2006correlations,kondor2014rich}. Our modeling reveals that complex organizations are generally not consistent with a power-law rank-frequency distribution of their components, 
and that the rate at which they are created and expand reflects trade-offs in the determination of self-similarity in the structure of the system. In many cases, these trade-offs lead to non-Zipfian distributions over a system's elements and self-similar behaviors that manifest themselves in cross-system data collapse under appropriately scaled variables. Self-similarity is a cornerstone of complex systems modeling \cite{Barenblatt}, and in this case allows for the unification of fundamentally distinct systems under a common theory of functional diversity.

\begin{center}
\begin{tcolorbox}[width=0.5\textwidth,colback={myblue!20},title={\small \textbf{Box 1: Glossary}},colbacktitle=mygreen!20,coltitle=black] 
   \small{
    \textbf{System size}: The number of agents/workers composing the system, denoted by $N$.\\
    \textbf{Function diversity}: The number of unique functions in a system, denoted by $D(N)$. In a city, this is the number of unique occupations. In a cell, this is the number of unique proteins.\\
    \textbf{Abundance distribution}: The set of unique functions and associated abundances encapsulated by $n_k(N)$, the number of functions of size $k$.\\
    \textbf{Rank-frequency distribution}: Upon assigning rank-order to the functions in a system, the rank-frequency distribution is a plot of size-normalized frequency vs rank. Eq.~\eqref{eq:r_def} shows how a function's rank, $r$, and its normalized frequency, $\tilde k(r) = k(r)/N$, are related. \\
    \textbf{Normalized rank-frequency distribution}: The analog of the rank-frequency distribution but for the \textit{normalized rank}, $\tilde r = r/D(N)$. \\
    \textbf{Zipf-Mandelbrot law}: Empirical observation of a power-law tail in a system's rank-frequency distribution. \\
    \textbf{Heaps' law}: Empirical finding of a power-law growth of diversity, i.e., $D\sim N^\beta,\; 0<\beta<1$.\\
    \textbf{Linear preferential attachment}: The probability to expand an existing function being \textit{linearly} proportional to the function's current size.\\
    \textbf{Yule-Simon model}: The seminal model of growing function diversity, with a constant probability of adding new functions, and linear preferential attachment to expand existing functions.\\
    \textbf{Self-similarity}: The finding that data from distinct systems, when appropriately rescaled or transformed, collapse onto overlapping curves or patterns that exhibit the same structure across different scales. \textit{Self-similarity} is also referred to as \textit{universality} in related literature \cite{youn2016scaling}.
    }
    \label{box:gloss}
\end{tcolorbox} 
\end{center}

\subsection*{Roadmap}
\noindent The main text reports the key mathematical formulae from the model and gives intuitive explanations for more complicated results; boxes contain more detailed calculations that are explained more intuitively in the main text; and the SI contains more elaborate calculations beyond the scope of the main text. In Section \ref{sec:model}, we introduce our model of function diversity and solve for the abundance distribution in the large $N$ limit. Then, in Sections \ref{sec:sublinpa} and \ref{sec:linpa}, we solve for the rank-frequency distributions in the case of sublinear and linear preferential attachment respectively. In Section \ref{sec:data}, we validate the model in the context of real data, and use the model to explain the occurrence of distinct qualitative phenomena across the data. Finally, in the discussion, we summarize the paper, discuss limitations, and propose future directions of the model. A summary of the key terms used in this manuscript is provided in Box 1.

\section*{Results}

\section{Abundance distributions and diversity scaling}\label{sec:model}
\noindent The model we use to describe the evolution of function diversity and system structure was first introduced in \cite{yang2025principles}, where a limited analytic treatment was given. The model is Yule-Simon-like, but it differs in two key respects: (i) the rate at which new functions are added is a decreasing function of the size of the system; and (ii) there is a generalized preferential attachment to join existing functions that can be linear or sub-linear. In other words, the probabilities, $p$ and $q_k$, are now generalized to have non-trivial dependence on $N$ and $k$. 

Consequently, we introduce
\begin{align}\label{eq:p_def}
    p(\theta,N)= \frac{p_0}{\sum_{k=1}^{k_{\mathrm{max}}} k^\theta n_k(N)}
\end{align}
as the probability of creating a new function, and 
\begin{align}\label{eq:q_def}
    q_k=\frac{k^\gamma}{\sum_{k=1}^{k_{\mathrm{max}}} k^\gamma n_k(N)}
\end{align}
as the probability of expanding a new function with $k$ members. The sums over $n_k$ in the denominators of these functions act to scale $p$ and $q_k$ as the system gets bigger. For $q_k$, the sum is a normalization factor such that $\sum_k q_k n_k = 1$. The upper limit of the sums, $k_\mathrm{max} = k_\mathrm{max}(N)$, is also dependent on $N$. The form of $q_k$ can be intuited by considering $\gamma=0$ or 1, which correspond to no preferential attachment and linear preferential attachment, respectively, following \cite{krapivsky2001organization}. The functional form of $p(\theta,N)$ is the simplest function that: (i) is a function of all functions in the organization (i.e., $n_k(N)\;\forall\; k$); (ii) flexibly accounts for a sublinear growth of diversity across complex systems \cite{yang2025principles,okie2025equilibrium}; and (iii) allows for an analytically tractable model. If one member is added to the organization per unit time, then the master equation that describes the growth of $n_k$ is the same as Eq.~\eqref{eq:genME} with the newly defined $p$ and $q_k$ from Eqs.~\eqref{eq:p_def} and \eqref{eq:q_def}. 

In the special cases of $\theta=\gamma=0$ and $\theta=\gamma=1$ one can solve this set of equations explicitly (in some cases exactly), and we show these cases in Sec.~\ref{sec:exact}. Further references to Eq.~\eqref{eq:genME} use $p$ and $q_k$ from Eqs.~\eqref{eq:p_def} and \eqref{eq:q_def}. Rather than solving Eq.~\eqref{eq:genME} on a case-by-case basis, useful asymptotic formulae can be extracted for general $\theta$ and $\gamma$ for large $k$ and $N$. Below we present a concise asymptotic solution to Eq.~\eqref{eq:genME}. Please refer to Section \ref{sec:leading_deriv} for more details on the calculations below. 


\begin{figure}[ht]
    \includegraphics[width=.5\textwidth]{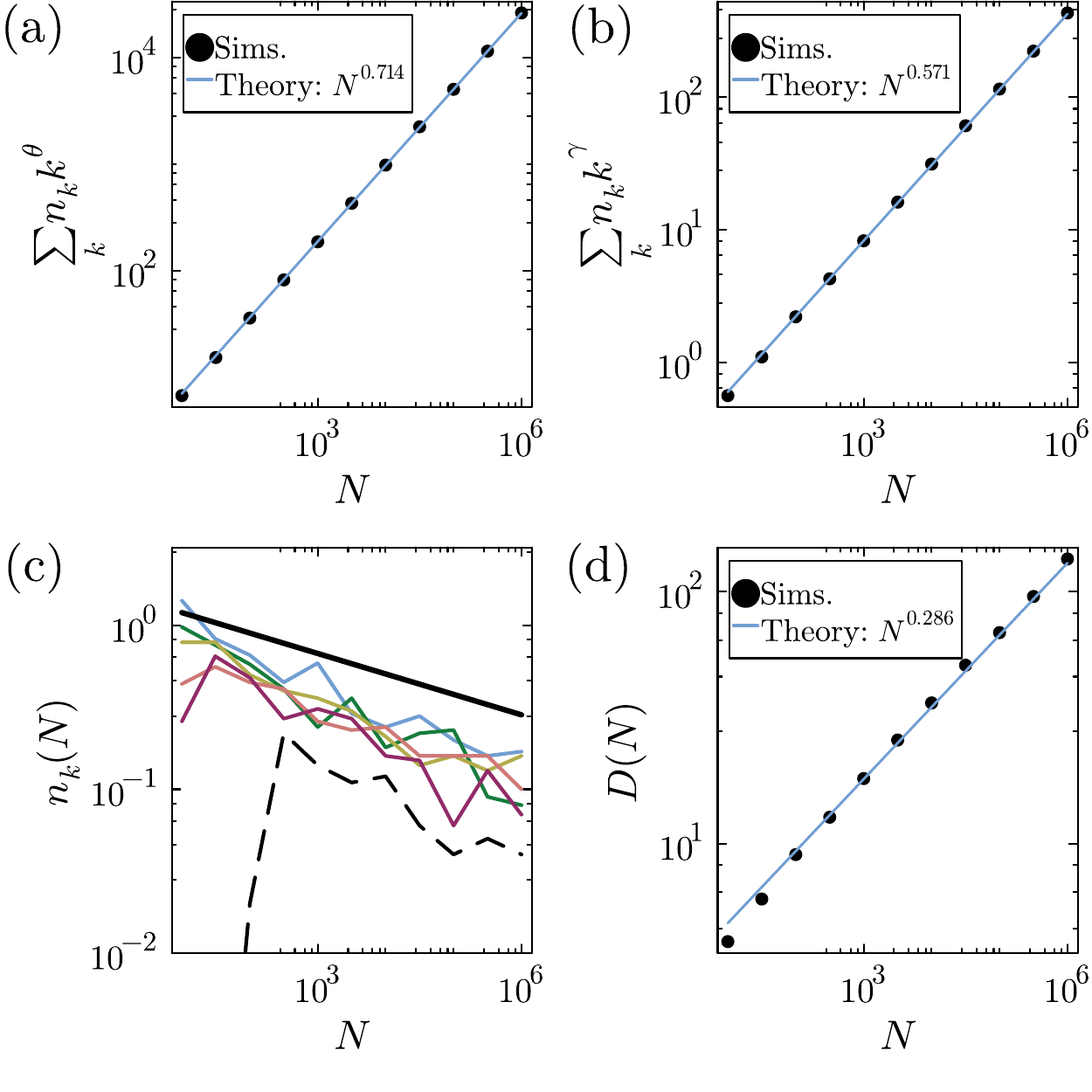}
    \caption{\textbf{For $\theta = 0.6$ and $\gamma = 0.4$ we verify the predicted asymptotic scaling laws, plotted on log-log axes.}
    Each simulated data point is an average over $10^2$ simulations. For (a), (b) and (d) the black dots represent the simulations and the blue line represents the theory. In (c): the solid black line is the prediction $N^{-0.125}$ from Eq.~\eqref{eq:nk-first}; the blue line is $n_1(N)$ from the simulations; the green line is $n_2(N)$; the yellow line is $n_3(N)$; the orange line is $n_4(N)$; the red line is $n_5(N)$; and the black dashed line is $n_{30}(N)$. There is a required size until which the scaling holds for each $k$ which gets larger as $k$ gets larger, most noticeable in the case $n_{30}(N)$.}
    \label{fig:conf-analytics}
\end{figure}

The first part of the analytics involves finding asymptotic relationships for the sums over $n_k$ from the definitions of $p$ and $q_k$ in Eqs.~\eqref{eq:p_def} and \eqref{eq:q_def}. This is possible by summing Eq.~\eqref{eq:genME} over $k$, and utilizing the relationships $N = \sum_k kn_k$, $D(N) = \sum_k n_k$, and $D(N) = \int p(\theta,N) \mathrm{d}N$. The analysis in Sec.~\ref{sec:ansatz} reveals that
\begin{align}\label{eq:sum-scaling}
    \sum_k k^x n_k(N) \sim N^{\frac{x-1}{2-\theta}+1}.
\end{align}
Although $p$ and $q_k$ follow relatively complex definitions, they exhibit simple asymptotic relationships with $N$ via Eq.~\eqref{eq:sum-scaling}, $\sum_k k^\theta n_k(N)\sim N^\nu$ and $\sum_k k^\gamma n_k(N)\sim N^\mu$, in which scaling exponents $\mu$ and $\nu$ are given by
\begin{align}
    \nu = \frac{1}{2-\theta},\quad \mu = \frac{1+\gamma-\theta}{2-\theta}.
\end{align}
In Fig.~\ref{fig:conf-analytics}(a)-(b), we verify these scaling relationships for $\theta = 0.6$ and $\gamma = 0.4$. Using the asymptotics derived above, one can solve Eq.~\eqref{eq:genME} in the case of sublinear preferential attachment to find
\begin{align}\label{eq:nk-first}
    n_k(N) \sim k^{-\gamma} N^{\mu-\nu},\; \gamma<1.
\end{align}
That is, \textit{$n_k$ scales as a power law in both $N$ and $k$}.

In the case of linear preferential attachment, the solution is slightly different, due to different analytic approximations used in Sec.~\ref{sec:ansatz}. The result is
\begin{align}\label{eq:nk-lpa}
    n_k(N) \sim  k^{\nu-2} N^{1-\nu},\; \gamma=1.
\end{align}
Therefore, both linear and sub-linear preferential attachment admit similar power-law dependence in $N$ and $k$, but for linear preferential attachment, the exponent of $k$ varies in response to changes in $\theta$. 

The scaling dependence of $N$ in Eqs.~\eqref{eq:nk-first} and \eqref{eq:nk-lpa} tells us how functions of a fixed size $k$ change in abundance as system size increases. The scaling in Eqs.~\eqref{eq:nk-first} and \eqref{eq:nk-lpa} also states that if $\gamma=\theta$ (in which case $\mu = \nu$) then $n_k$ becomes independent of $N$, which follows from the substitution $\theta\to\gamma$ in Eq.~\eqref{eq:p_def}, and the subsequent realization that the probability of adding to a function of any size then admits the same $N$ dependence. Having $n_k$ independent of $N$ for $\gamma=\theta$ exhibits the profound trade-offs between the strength of preferential attachment and the rate at which new functions are added. We verify the scaling of $n_k(N)$ in Eq.~\eqref{eq:nk-first} for various values of $k$ and $N$, $\theta = 0.6$ and $\gamma = 0.4$ in Fig.~\ref{fig:conf-analytics}(c).

Our analytic framework allows for easy identification of the growth of a system's diversity. The diversity is the cumulative sum of the number of new functions added across all time steps, plus the number of functions initially present. Taking $n_k(1)=\delta_{k,1}$, this can mathematically be written as
\begin{align}\label{eq:d_scaling}
    D(N)= \int^N_1 p(\theta,N')\mathrm{d}N'\sim 
    \begin{cases}
        N^{\frac{1-\theta}{2-\theta}}, &\theta\in [0,1),\\
        \ln(N), &\theta = 1,
    \end{cases}
\end{align}
following Eqs.~\eqref{eq:p_def} and \eqref{eq:sum-scaling}.
In Sec.~\ref{sec:minimal-set}, we derive how the functional form of $D(N)$ changes when there is initially more than a single agent, i.e., $n_k(1)\neq\delta_{k,1}$. 

In the following sections, we consider the regimes of sub-linear and linear preferential attachment separately, and show how the approach from sublinear to linear preferential attachment can be seen as a phase transition. We show that different types of self-similarity in rank-frequency distributions are admitted to systems with linear versus sub-linear preferential attachment. These self-similarities allow one to distinguish between different phases of the model's $(\theta,\gamma)$ parameter space. These rank-frequency distributions then allow us to explain complex features of empirical data in Section \ref{sec:data}.


\section{Rank-frequency distributions}
\subsection{Sub-linear preferential attachment}\label{sec:sublinpa}
\noindent For sub-linear preferential attachment, the number of functions of size $k$ scales as in Eq.~\eqref{eq:nk-first}. Here, we calculate two quantities that are reflective of the dynamics of the growth of a system with sub-linear preferential attachment: (1) the scaling of the largest expected function size, and (2) the functional form of the rank-frequency distributions over system components. The scaling of the largest function size allows for calculation of the rank-frequency distributions which provide key insights into a complex system's structure, and a system's potentially universal properties. 


To determine the scaling of the largest expected function size, consider an alternative, but equally valid, definition of diversity that is distinct from Eq.~\eqref{eq:d_scaling},
\begin{align}\label{eq:div2}
    D(N) = \sum_{k=1}^{k_{\mathrm{max}}(N)}n_k.
\end{align}
This states that the diversity is the sum of the number of functions of any size. 
Equating Eqs.~\eqref{eq:nk-first} and \eqref{eq:div2}, and using Eq.~\eqref{eq:nk-first}, leads to $k_{\mathrm{max}}(N)\sim N^{1/(2-\theta)}$. 
Surprisingly, the scaling of the largest function does not depend on the strength of preferential attachment in the regime of sublinear preferential attachment. For example, this scaling law indicates that if the size of the largest city is only dependent on a country's size, then city growth subscribes to sublinear preferential attachment.

\begin{center}
\begin{tcolorbox}[width=0.5\textwidth,colback={myblue!20},title={\small \textbf{Box 2: Asymptotic rank-frequency distributions for sublinear preferential attachment}},colbacktitle=mygreen!20,coltitle=black] 
   \small{
    Both $k_{\mathrm{max}}(N)$ and $D(N)$ have multiplicative pre-factors in their scaling relationships with $r$ and $N$. Here we neglect these for brevity, with a main focus on how the rank-frequency distributions functionally depend on $r$ and $N$. 
    
    Using Eq.~\eqref{eq:nk-first}, and computing Eq.~\eqref{eq:r_def}, we find
    \begin{align}\label{eq:r_sublin}
    r\sim
    \begin{cases}
        \frac{D(N)}{1-\gamma}\left[ 1-\left(\frac{k(r)}{k_{\mathrm{max}}(N)}\right)^{1-\gamma} \right], &\gamma<1,\theta<1,\\
        \frac{1}{1-\gamma}\left[ 1-N^{\gamma-1}k(r)^{1-\gamma} \right], & \gamma<1,\theta=1.
    \end{cases}
\end{align}
We define $\tilde{k}(r)=k(r)/N$ as the normalized frequency of the function with rank $r$. Rearrangement of Eq.~\eqref{eq:r_sublin} gives
\begin{align}\label{eq:sublin-kr1}
    \tilde{k}(r) \sim
    \begin{cases}
        \left[D(N)-(1-\gamma)r\right]^{\frac{1}{1-\gamma}}, &\gamma<1,\theta<1,\\
        \left[ 1-r(1-\gamma) \right]^{\frac{1}{1-\gamma}}, & \gamma<1,\theta=1.
    \end{cases}
\end{align}
Eqs.~\eqref{eq:sublin-kr1} predict that one should observe rank-frequency distributions independent of $N$ when $\gamma<1$ and $\theta=1$. This independence is not observed for $\theta<1$.

Defining $\tilde{r} = r/D(N)$ as the normalized rank, one finds that
\begin{align}\label{eq:sublin-kr2}
    \tilde{k}(\tilde{r}) \sim
    \begin{cases}
        [1-(1-\gamma)\Tilde{r}]^{\frac{1}{1-\gamma}}, &\gamma<1,\theta<1,\\
        \left[ 1-\tilde{r}(1-\gamma)\ln(N) \right]^{\frac{1}{1-\gamma}}, & \gamma<1,\theta=1,
    \end{cases}
\end{align}
in which $\ln(N)$ comes from $D(N)$ evaluated at $\theta=1$.
Eq.~\eqref{eq:sublin-kr2} predicts $\tilde{k}(\tilde{r})$ is independent of $N$ for $\theta<1$. The quantity $\tilde{k}(\tilde{r})$ is related to $\tilde{k}(r)$ via the Jacobian transformation $\tilde k(\tilde r) = |\mathrm{d}r/\mathrm{d} \tilde r| \tilde k(r)$ such that $\int_0^1 \mathrm{d}\tilde r \tilde k(\tilde r) = 1$.
   }
\end{tcolorbox} 
\end{center}

To derive the rank-frequency distribution, we use the definition of a function's rank, $r$, given an abundance, $k(r)$,
\begin{align}\label{eq:r_def}
    r = 1+\int_{k(r)}^{k_{\mathrm{max}}(N)}n_{k'}(N)\mathrm{d}k',
\end{align}
which states that the rank is equal to the number of functions with an abundance greater than $k(r)$, plus 1. In Box 2, using the derived scaling for $k_{\mathrm{max}}$ and Eq.~\ref{eq:nk-first}, we derive the large $N$ asymptotics for both the \textit{rank-frequency}, $\tilde k(r)$, and \textit{normalized rank-frequency}, $\tilde k(\tilde r)$, distributions (see Box 1 for definitions). Briefly, the rank-frequency distribution assesses how a function's normalized abundance, $k/N$, depends on its rank of abundance, $r$, while the normalized rank-frequency distribution assesses how a function's normalized abundance depends on its normalized rank, $\tilde r = r/D(N)$. Variables with a tilde denote normalized quantities $\in [0,1]$. 

\begin{figure}[ht]
    \includegraphics[width=.5
\textwidth]{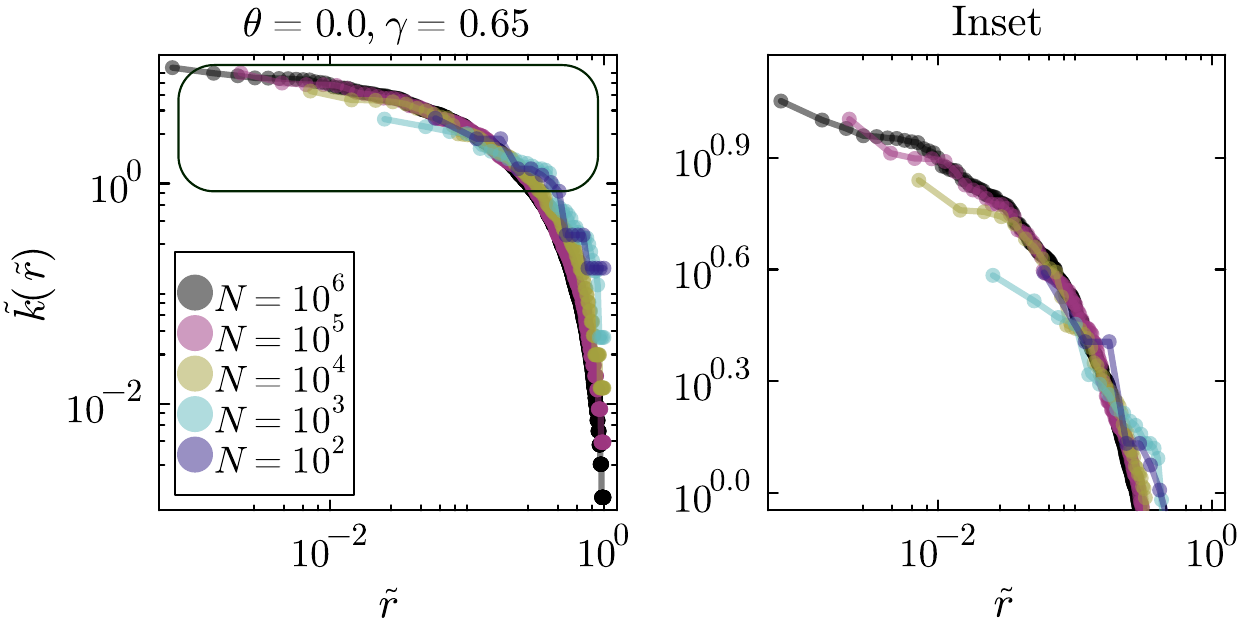}
    \caption{\textbf{Simulations showing the emergence of self-similarity in the normalized rank-frequency distribution, predicted by Eq.~\eqref{eq:sublin-kr2}, across values of $N$ when $\gamma=0.65$ and $\theta=0$.} The area under each curve is 1 from the definition of $\tilde k (\tilde r)$. The inset shows increasing agreement with the largest system as $N$ increases further into the asymptotic regime.}
    \label{fig:univ-sims}
\end{figure}

The analytics in Box 2 reveal that sublinear preferential attachment leads to: (i) \textit{rank-frequency distributions that are not Zipfian}; (ii) rank-frequency distributions that are not self-similar---i.e., independent of system size, $N$---unless diversity growth is logarithmic; and (iii) self-similar normalized rank-frequency distributions (see simulation verification in Fig.~\ref{fig:univ-sims})---unless diversity growth is logarithmic, in which case the distributions are \textit{almost self-similar} (due to the slow change of $\ln N$). In Fig.~\ref{fig:non-naive} (row 1 and 2), we show that theoretical predictions (i)-(iii) also hold when one does not have an initial condition in which $n_k(1)=\delta_{k,1}$.

\begin{figure}[ht]
    \includegraphics[width=.5\textwidth]{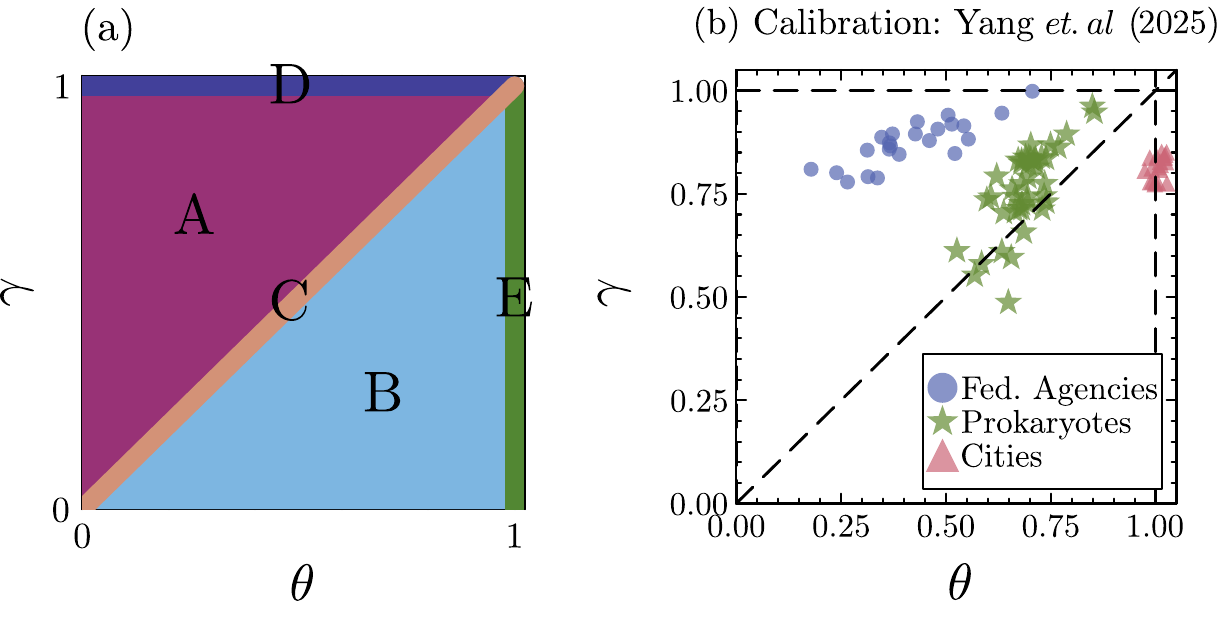}
    \caption{\textbf{Phase diagram of the behaviors expressed by the model in the range $0\leq\theta\leq 1$ and $0\leq\gamma\leq1$.} (a) Phases in the parameter space. For the characteristics of each phase see Table \ref{tab:phases}. (b) Placement of real systems in the phase space, reproduced from \cite{yang2025principles}. Each individual point represents a calibrated value of $(\theta,\gamma)$ for a single system, i.e., an individual city or protein abundance distribution. The calibration procedure used in \cite{yang2025principles} minimized the distance between the rank-frequency distributions from the data and those from the model in order to choose the optimal values of $\theta$ and $\gamma$.}
    \label{fig:phase-d}
\end{figure}

These results suggest a phase separation between logarithmic and power-law diversity growth in the regime of sublinear preferential attachment. Fig.~\ref{fig:phase-d}(a) and Table \ref{tab:phases}, summarize these results in a phase diagram, which indicates several differences in $D(N)$, $n_k$, $\tilde k(r)$ and $\tilde k(\tilde r)$ across the $(\theta,\gamma)$ parameter space. Table \ref{tab:phases} explains the properties of the labeled phases in Fig.~\ref{fig:phase-d}(a).

\begin{table*}[t]
\centering
\renewcommand{\arraystretch}{1.4} 
\begin{tabular}{|c|c|c|c|c|c|}
\hline
\textbf{Phase} &
\(D(N)\) \textbf{power law?} &
\(n_k(N) \;\) \textbf{dependence on} \(N\) &
\(\tilde{k}(r)\) \textbf{self-similar?} &
\(\tilde{k}(\tilde{r})\) \textbf{self-similar?} & $\tilde k (r)$ \textbf{Zipfian}? \\
\hline
A & \cmark & increasing & \xmark & \cmark & \xmark \\
B & \cmark & decreasing & \xmark & \cmark & \xmark \\
C & \cmark & stays the same & \xmark & \cmark & \xmark \\
D & \cmark & increasing & \cmark & \xmark & \cmark \\
E & \xmark & decreasing & \cmark & $-$ & \xmark \\
\hline
\end{tabular}
\caption{Table explaining the properties of the five phases from Fig.~\ref{fig:phase-d}. 
In the second column, phase E has $D(N)\sim \ln(N)$. 
In the fifth column, the ``--'' indicates that $\tilde{k}(\tilde{r})$ is almost self-similar with respect to increasing $N$ for phase E.}
\label{tab:phases}
\renewcommand{\arraystretch}{1.0}
\end{table*}

For sublinear preferential attachment with power-law diversity growth ($\theta<1$), the deviation from Zipf's law for rank-frequency distributions is subtle, yet stark. Although the distributions share a similar functional form to the Zipf-Mandelbrot law \cite{piantadosi2014zipf}, the form $(\beta-r)^\alpha$ for $\alpha >0$ (from Eq.~\eqref{eq:r_sublin}), is qualitatively different than $(\beta+r)^{-\alpha}$ (the Zipf-Manderbrot law) and does not admit a power-law tail. Notably, both $(\beta-r)^\alpha$ and $(\beta+r)^{-\alpha}$ can share very similar qualitative features across the $(\alpha,\beta)$ parameter space---both can show log-log space curvature. Therefore, our theory provides an important theoretical perspective to distinguish between processes that directly lead to power-law distributed components, and those that do not. This is especially useful in cases where data sparsity does not indicate clear power-law behaviors. 

Finally, sublinear preferential attachment, while still giving rise to the power-law scaling $n_k\sim k^{-\gamma}$ for the abundance distribution, does not give power-law scaling for the rank-frequency distribution. This is surprising in the light of previous studies \cite{zanette2005dynamics,loreto2016dynamics} in which power-law distributed abundance always gives rise to power-law distributed rank-frequency distributions. Therefore, our model provides a framework to describe systems with distinct scaling between their abundance and rank-frequency distributions.

\subsection{Linear preferential attachment}\label{sec:linpa}
\noindent In the regime of linear preferential attachment the number of functions of size $k$ scales as in Eq.~\eqref{eq:nk-lpa}. Following similar arguments as the previous section, one can show that the scaling of the largest function is now $k_{\mathrm{max}}(N)\sim N$, indicating that for linear preferential attachment, how new functions are added is relatively unimportant, and that a few large functions begin to dominate the system. The scaling $k_\mathrm{max}\sim N$ is exactly recovered in the special case solution of $\gamma=\theta=1$ in Section \ref{sec:exact1}.

The analytics in Box 3 reveal that linear preferential attachment is qualitatively distinct from sublinear preferential attachment. In particular: (i) \textit{the rank-frequency distributions are always Zipfian}, unless diversity growth is logarithmic, in which case they exhibit an exponential decay; (ii) the rank-frequency distributions are always self-similar; (iii) the normalized rank-frequency distributions are never self-similar. In Fig.~\ref{fig:non-naive} (row 3 and 4) we show that these theoretical predictions hold when one does not have an initial condition in which $n_k(1)=\delta_{k,1}$.

\begin{center}
\begin{tcolorbox}[width=0.5\textwidth,colback={myblue!20},title={\small \textbf{Box 3: Asymptotic rank-frequency distributions for linear preferential attachment}},colbacktitle=mygreen!20,coltitle=black] 
   \small{
    For the rank, using Eqs.~\eqref{eq:nk-lpa} and \eqref{eq:r_def}, we can now write
    \begin{align}\nonumber
        r \sim N^{1-\nu} \int_{k(r)}^N k^{\nu-2}\mathrm{d}k.
    \end{align}
    Evaluating the integral, ignoring asymptotic pre-factors, and rearranging, we arrive at
    \begin{align}\label{eq:linPA_k1}
        \tilde{k}(r) \sim
        \begin{cases}
            (r(1-\nu)+\nu)^{\frac{1}{\nu-1}}, &\gamma=1,\theta<1,\\
            e^{-r}, &\gamma=\theta=1.
        \end{cases}
    \end{align}
    With respect to the normalized rank we find
    \begin{align}\label{eq:linPA_k2}
        \tilde{k}(\tilde{r}) \sim
        \begin{cases}
            (\tilde{r}(1-\nu)D(N)+\nu)^{\frac{1}{\nu-1}}, &\gamma=1,\theta<1,\\
            N^{-\tilde{r}}, &\gamma=\theta=1.
        \end{cases}
    \end{align}
    These results imply that when there is linear preferential attachment, $\tilde k(r)$ is always self-similar regardless of how new functions are added to the system. On the other hand, we observe a lack of self-similarity in $\tilde{k}(\tilde{r})$.
   }
\end{tcolorbox} 
\end{center}

In addition to the phases already identified for the case of sublinear preferential attachment, the approach $\gamma\to 1^-$ constitutes a phase transition in our model. This is represented as phase D in Fig.~\ref{fig:phase-d}. Phase D differs in many respects as compared to the other four phases, as is clear from Table \ref{tab:phases}. Most pertinently, unlike sublinear preferential for $\theta<1$, the rank-frequency distributions are Zipfian for linear preferential attachment, in addition to being self-similar. This aligns with previous predictions in \cite[Eq.~(23)]{zanette2005dynamics} for a special case of our model. The opposite is true of the normalized rank-frequency distributions as the distributions are not self-similar, unlike for sublinear preferential attachment. Only in the case of sublinear preferential attachment with logarithmically growing diversity are there similarities to the qualitative properties of self-similarity for linear preferential attachment.



\section{Empirical validation and prediction}\label{sec:data}
\noindent To assess the realism of our model of function diversity, we now assess the model's ability to predict aspects of empirical data across cells, federal agencies and urban areas (data that was previously introduced in \cite{youn2016scaling,yang2025principles}). In the analysis below, we show that our model predicts the qualitative properties of observed distributions across many modalities. 

The data comprises:
\begin{enumerate}
    \item[1.] Protein abundance distributions in prokaryotic organisms, normalized for a single cell \cite{arroyo2024algorithm}. This data spans 47 different species of prokaryotic cell.
    \item[2.] Federal agency employee data taken from FedScope Employment Cube provided by the US Office of Personnel Management (OPM). This data spans 125 US government federal agencies.
    \item[3.] City occupation data taken from the US Bureau of Labor Statistics (BLS). This data spans 422 cities.
\end{enumerate}
For more details on the data, please see \cite[Sec.~S1]{yang2025principles}

\begin{figure}[ht]
    \includegraphics[width=.45\textwidth]{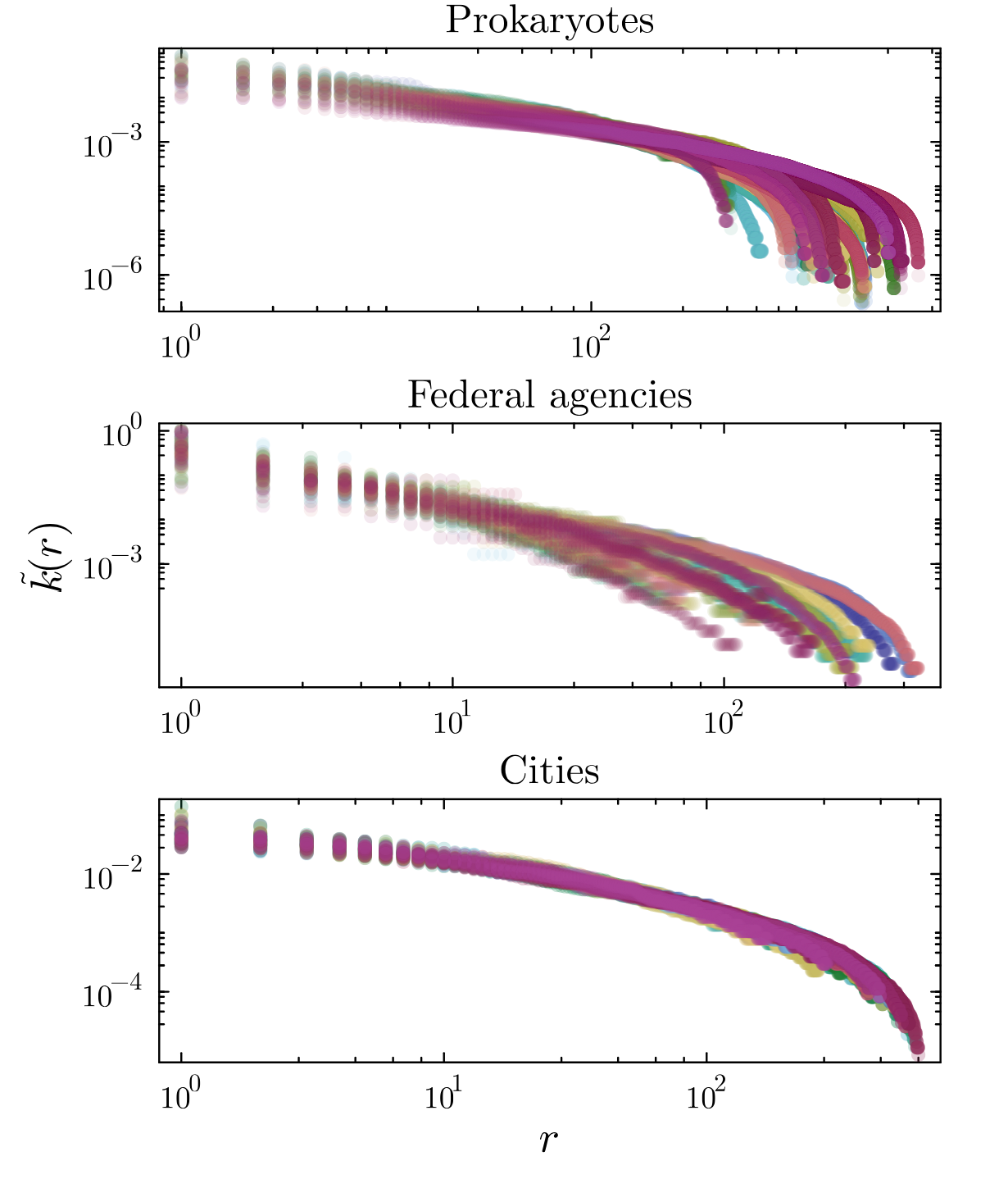}
    \caption{\textbf{Rank-frequency distributions, $\Tilde{k}(r)$, of proteins in bacteria, employees in federal agencies, and workers in cities.} Each color represents a different system, e.g., a specific city or protein abundance distribution. Protein rank-frequency distributions in prokaryotes and employee rank-frequency distributions in federal agencies are not self-similar. However, rank-frequency curves for employees across cities do subscribe to self-similarity, as all distributions collapse onto the same curve. This is predicted by Eq.~\eqref{eq:sublin-kr1} for $\theta=1$.}
    \label{fig:rank-freq}
\end{figure}

\begin{table*}[t]
\centering
\renewcommand{\arraystretch}{1.4} 
\begin{tabular}{|l|c|c|c|c|c|}
\hline
\textbf{System} &
\(D(N)\) \textbf{power law?} &
\(n_k(N) \;\{\uparrow,-,\downarrow\}\) \textbf{with } \(N\)? &
\(\tilde{k}(r)\) \textbf{self-similar?} &
\(\tilde{k}(\tilde{r})\) \textbf{self-similar?} & \textbf{Phase} \\
\hline
Federal agencies & \cmark & $\uparrow$ & \xmark & \cmark & A \\
Prokaryotes & \cmark & $\uparrow$& \xmark & \cmark & A \\
Cities & \xmark & $\downarrow$ & \cmark & ?/\cmark & E\\
\hline
\end{tabular}
\caption{Table placing the three empirical systems into their respective phases. For empirical identification of the phases, the properties of Table \ref{tab:phases} were used and compared to Figs.~\ref{fig:rank-freq}, \ref{fig:rank-freq-univ} and \ref{fig:si-nk}. The empirical phases identification based on Table \ref{tab:phases} is in exact agreement with the calibration results provided in Fig.~\ref{fig:phase-d}(b). }
\label{tab:empirics}
\renewcommand{\arraystretch}{1.0}
\end{table*}

\begin{figure}[ht]
    \includegraphics[width=.45\textwidth]{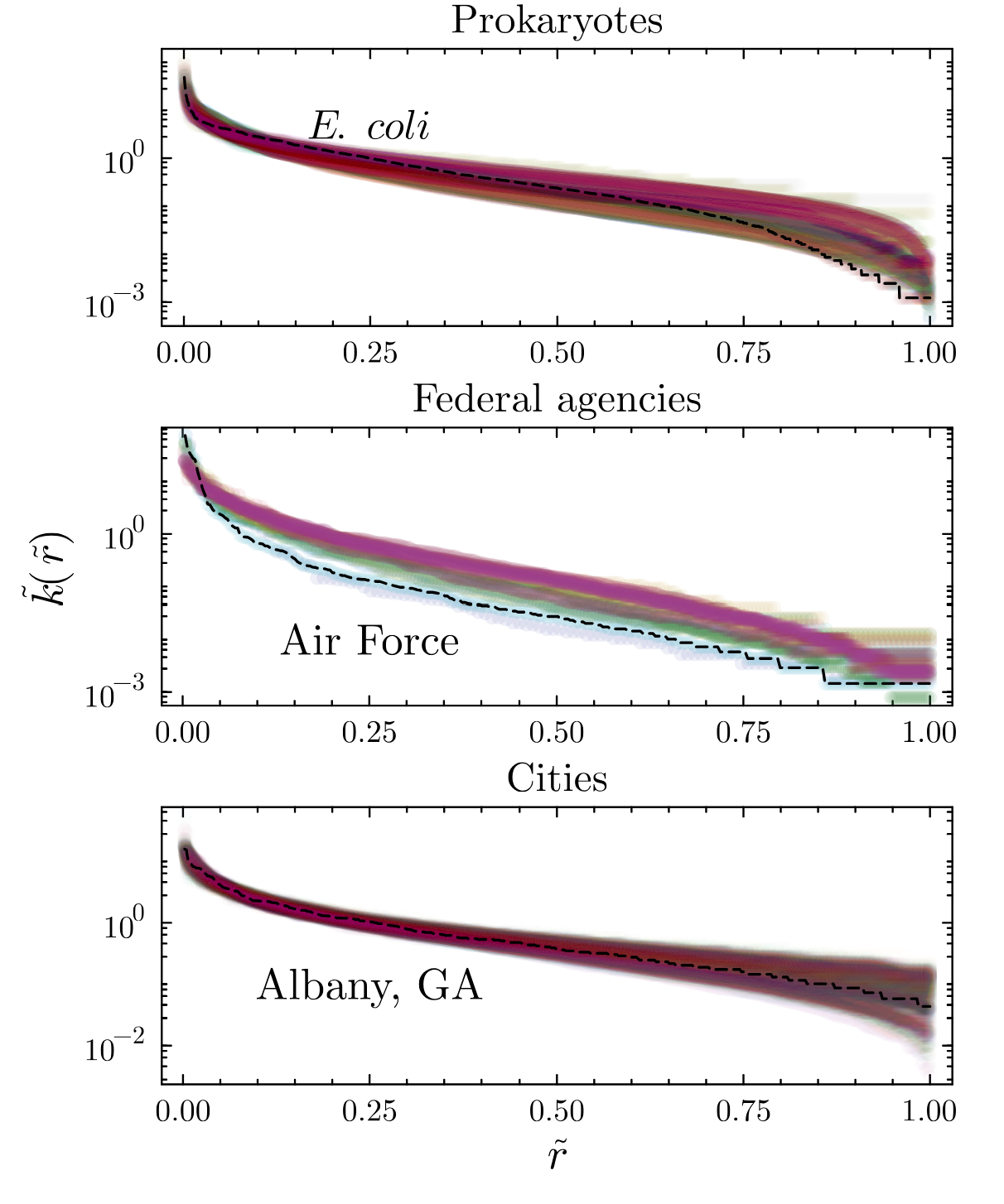}
    \caption{\textbf{Normalized rank-frequency distributions, $\tilde{k}(\tilde{r})$, of proteins in prokaryotes, employees in federal agencies, and workers in cities.} Each $x$-axis is re-scaled such that the rank is represented on $[0,1]$ for all systems in that class. The area under each curve is 1.  Despite the noise in predictions of $\gamma$ and $\theta$ across different prokaryotes and federal agencies (see Fig.~\ref{fig:phase-d}(b)), we find that the rank-frequency curves share a self-similarity for the rank-normalized rank-frequency distributions. On the other hand, cities—which share broadly the same values of $\gamma$ and $\theta$—are almost self-similar aside from values of $\tilde r\lesssim 1$. These results are predicted by Eqs.~\eqref{eq:sublin-kr2}. Due to sample noise, only federal agencies with over 1000 members are included in the figure. All cities and prokaryotes in the data are included. Dashed lines highlight individual examples from the respective data.}
    \label{fig:rank-freq-univ}
\end{figure}

In Fig.~\ref{fig:rank-freq} we plot the normalized rank-frequency distributions, $\Tilde{k}(r)$, for proteins in prokaryotes, and employees making up federal agencies and cities. We plot the same systems in Fig.~\ref{fig:rank-freq-univ}, but now for the normalized rank-frequency distributions wherein rank is also normalized, i.e., $\tilde{k}(\tilde{r})$. 

Several points become clear from the data. First, none of the rank-frequency distributions in Fig.~\ref{fig:rank-freq} is Zipfian in nature---there are no straight lines in the log-log space plots of their rank-frequency curves which would indicate clear power-law behaviors. Second, for the rank-frequency distributions in Fig.~\ref{fig:rank-freq} only the cities exhibit a collapse to a single curve indicative of self-similarity, while for prokaryotes and federal agencies, such self-similarity is not observed. Third, across all three systems in Fig.~\ref{fig:rank-freq-univ}, approximate self-similarity in the normalized rank-frequency distributions is seemingly observed.

Our analytic predictions agree with a previous computational approach that calibrated the data to the model \cite{yang2025principles}. Here, we show the calibrated parameter values from \cite{yang2025principles} in Fig.~\ref{fig:phase-d}(b), which clearly shows that: (i) $\gamma<1$ for all systems; (ii) for prokaryotes and federal agencies, diversity increases as a power law in $N$ ($\theta<1$); and (iii) for cities diversity growth is logarithmic across all cities ($\theta=1$). Following Eq.~\eqref{eq:sublin-kr1}, this confirms that the rank-frequency distributions are not Zipfian. The scatter of the data points in Fig.~\ref{fig:phase-d}(b) shows cities to be uniformly similar in type---with similar values of $\theta$ and $\gamma$---whereas prokaryotes and federal agencies differed quite significantly from one another in terms of $\theta$ and $\gamma$ values. That is, different cities create new functions and expand existing functions in very similar ways, whereas different prokaryotes and federal agencies do so in slightly different ways.

Our results shine new light on the results of \cite{yang2025principles}. By looking at the self-similarity present in each system, one can easily identify which phase of the $(\theta,\gamma)$ parameter space each system occupies (see Fig.~\ref{fig:phase-d}(a)). Since prokaryotes and federal agencies have self-similarity in the normalized rank-frequency distributions, but not the standard rank-frequency distributions, it becomes clear from Eqs.~\eqref{eq:sublin-kr1}--\eqref{eq:sublin-kr2} that they occupy the regime $\theta<\gamma<1$ (phase A in Fig.~\ref{fig:phase-d}(a)). This corresponds with the calibrations of $\theta$ and $\gamma$ in Fig.~\ref{fig:phase-d}(b). The noise in their self-similar behavior in Fig.~\ref{fig:rank-freq-univ} can then be seen through the slight variation in the values of $\theta$ and $\gamma$ in Fig.~\ref{fig:phase-d}(b). For the cities, the self-similarity in their rank-frequency distributions, and \textit{almost} self-similarity in their normalized rank-frequency distributions, rules out the mechanism of linear preferential attachment (which does not predict self-similarity of $\tilde{k}(\tilde{r})$), indicating that cities must occupy the regime of $\theta=1$ and $\gamma<1$ (phase E in Fig.~\ref{fig:phase-d} and Table \ref{tab:phases}). In Table \ref{tab:empirics} we have summarized the qualitative properties of each system, leading to a prediction for the phase which is confirmed in Fig.~\ref{fig:phase-d}(b).

Our model of functional diversity goes beyond context-dependent mechanisms of creating new functions and expanding existing functions to describe the structure of organizations across biology and society. However, the model does not include path dependence, function obsolescence, and non-tree-like growth, which could be reasonably hypothesized to be important in the formation of an organization's structure. For example, businesses die in cities. New federal agencies are often born from scratch in a top-down manner and don't undergo continuous bottom-up evolution to refine their structures (for example, the Office of the Department of Homeland Security, which had around 170,000 employees when it was established in 2002 \cite{bush_whitehouse_2002}). Protein abundance distributions in cells have a highly path-dependent evolutionary origin. That our simple model can capture the qualitative behaviors of such a diverse array of data may seem surprising in these contexts. 
Given the model's success, we can make further theory-driven predictions and assess the model's ability to describe finer aspects of the data.

\begin{figure}[ht]
    \includegraphics[width=.4\textwidth]{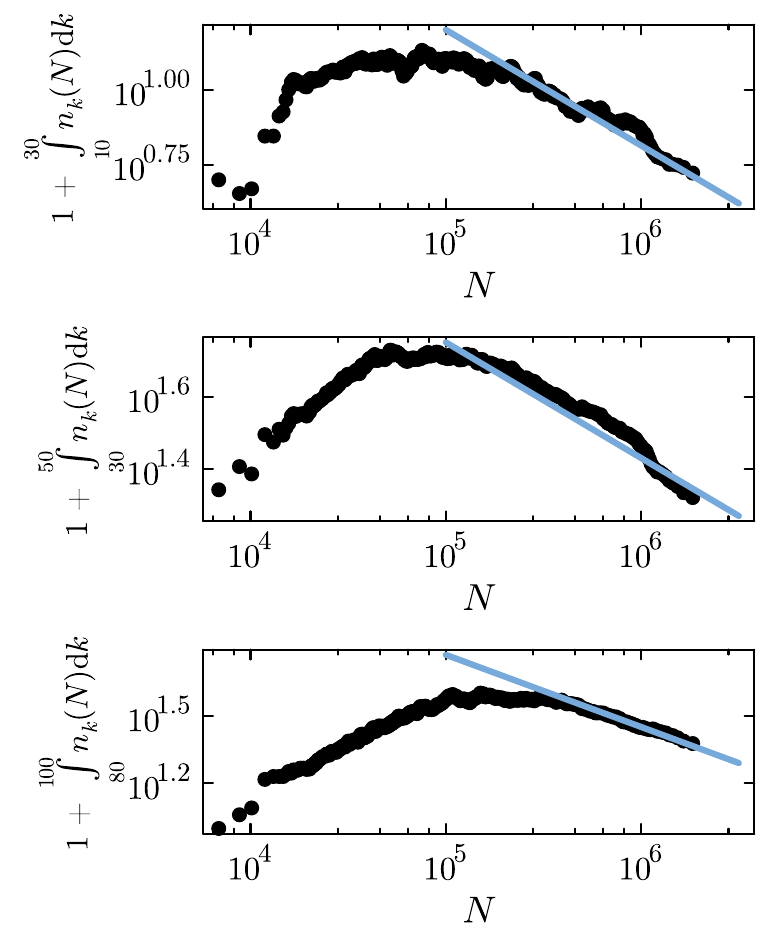}
    \caption{\textbf{Scaling of $n_k$ versus $N$ for the cities.} We integrate $n_k$ over a range of $k$ due to the sparsity of the data. Black dots show the running average of the integrated range of $n_k$ over 30 cities. Blue lines in each case show $N^{-0.32}$, whose value was chosen to provide the best fit for the tail of the top plot. This scaling is predicted by combined knowledge that for cities $\gamma<\theta$ \cite{yang2025principles} and Eq.~\eqref{eq:nk-first}. As observed in Fig.~\ref{fig:conf-analytics}(c), there is an initial size of city required for the predicted scaling to be observed for a given value of $k$.}
    \label{fig:city-nk}
\end{figure}


To assess the model's ability to make theory-driven predictions, we now look at whether the regimes of the $(\theta,\gamma)$ parameter space explain how the number of functions of size $k$, $n_k$, scales with respect to system size $N$. The model makes two predictions for the change in $n_k$ with $N$ from Eq.~\eqref{eq:nk-first}: (i) the change is independent of the value of $k$, and (ii) that the sign of $\gamma-\theta$ dictates whether $n_k$ increases, decreases or stays the same as $N$ varies. For prokaryotes and federal agencies, which are in the $\gamma>\theta$ regime (phase A in Table \ref{tab:phases} and Fig.~\ref{fig:phase-d}(a)), it should be the case that $n_k(N)$ increases as a function of $N$ since $\gamma-\theta>0$. On the other hand, for cities we find that $\gamma-\theta<0$ (phase E in Table \ref{tab:phases} and Fig.~\ref{fig:phase-d}(a)), implying that $n_k(N)$ should decrease as a function of $N$. These predictions are exactly what is observed in the data. In Fig.~\ref{fig:city-nk} we show that $n_k(N)$ scales sub-linearly in $N$ for US cities. In the SI, we verify the predicted scaling of $n_k(N)$ across all three systems in Fig.~\ref{fig:si-nk}. Therein, it is clear that regardless of the value of $k$: (i) the scaling exponent for $N$ is roughly the same value; and (ii) that for federal agencies and prokaryotes, $n_k$ is an increasing function of $N$, and for cities it is a decreasing function of $N$ (shown by the blue lines in the bottom three rows of Fig.~\ref{fig:si-nk}). Note that due to the sparsity of data, the scaling for $n_k(N)$ against $N$ was evaluated by integrating over a range of $k$ values. Therefore, our simple model, which captures only the processes of diversification and existing function expansion, provides a parsimonious mechanism to predict the future evolution of an organization as it increases in size.

Our model provides intuitive explanations of unintuitive qualitative patterns seen in the data. We have shown for cities that novel functional diversity is generated at a relatively slow rate. As these functions grow in size, the number of functions of size $k$---for those functions that have reached asymptotic scaling---will decrease as the system size increases. Therefore, large functions that have not yet reached asymptotic scaling pick up the system's excess mass. This means that an increasingly large fraction of a city's mass becomes concentrated in functions of a large size, via a mechanism that is distinct from preferential attachment. 

On the other hand, for federal agencies and prokaryotic cells, the number of functions of size $k$ is constantly increasing, and functional diversity also grows at a much faster rate than for cities. This indicates that functions of small sizes are always present, and asymptotically, the ratio of the numbers of functions of a given size remains constant. This is exemplified by the universality of the normalized rank-frequency distributions. Therefore, unlike cities which eventually have many functions of large sizes---and increasingly few functions of small size---it is always important for federal agencies and prokaryotes to have a diverse set of elements, many of which are not highly abundant.

\section*{Discussion}\label{sec:discussion}
\noindent We introduce a simple generative model to explain structural diversity across complex systems by mapping them into a parameter space defined by two fundamental processes: diversification—the creation of new categories—and preferential attachment—the accumulation of size or frequency within existing categories. These two mechanisms are among the fundamental principles governing the growth of complex systems and have been widely studied across disciplines, including linguistics, ecology, network science, and urban studies \cite{yang2025principles, zipf2016human, simon1955class}. 

What the model suggests is that a shared generative principle can manifest differently across domains—not because the mechanism changes, but because the conditions under which it operates vary. The implication is both strong and predictive: with just two parameters, the model can determine whether a system tends toward Zipfian or non-Zipfian distributions. In particular, systems governed by sublinear preferential attachment tend to deviate from Zipf’s law, as observed for a variety of empirical data in our study. Yet, despite Zipfian behavior emerging only within a narrow slice of this parameter space, its recurring presence across biological and social systems calls for the question: why do so many real-world systems fall within this specific parameter space? This is an open and compelling question for future research.

In the biological and urban scaling literature \cite{west1997general,kempes2011predicting,hamilton2007nonlinear,stier2023urban,youn2016scaling,bettencourt2013origins}, self-similar structures and power laws often highlight system characteristics that apply at every scale \cite{Barenblatt}. This is often interpreted as evidence for general principles of organization within a complex system. For example, Kleiber's law states that $B\sim M^{3/4}$, where $B$ is basal metabolic rate and $M$ is an organism's mass \cite{kleiber1947body}---a law which results from the minimization of energy dissipated by a circulatory system \cite{west1997general}. Our observation of hidden self-similar behaviors across the three systems studied herein paves the way for future hypotheses regarding the general principles that these behaviors derive from. In particular, the self-similarity of normalized rank-frequency distributions implies that complex systems evolve such that a single or small set of functions will not dominate over other functions---in addition to a requirement that the relative abundances of different functional components remain constant. However, the rank-frequency distributions of cities are also self-similar, suggesting that cities may be governed by organizing principles distinct from those of cells and federal agencies. 
Understanding how diversity patterns and resultant universalities are connected to the known constraints of prokaryotes and cities, which lead to distinctly different scaling relationships \cite{gastner2006optimal,bettencourt2007growth,stier2023urban,ritchie2023metabolic}, is a critical area of future work.


Limitations of our model arise because we do not include several processes. First, our model neglects the removal of agents or functions, which occurs in reality (e.g., when a colleague is fired, or when an organism evolves to a smaller size or loses genes). However, function obsolescence is a complex problem, and it is not simply function creation and growth in reverse \cite{saavedra2008asymmetric}. Realistic implementations of function obsolescence are not simple \cite{lee2025synthesis}. Second, our model considers $\theta$ and $\gamma$ to be fixed parameters that do not change in time---all the variation in the parameters being cross-sectional rather than longitudinal. Given results showing that cross-sectional and longitudinal scaling exponents for the size of city functional categories (e.g., lawyers, teachers, and miners) differ from each other \cite{hong2020universal}, this motivates considering time-varying $\theta$ and $\gamma$ in future work. Finally, we do not consider higher-order aspects of diversity that have been explored in other studies \cite{di2025dynamics}, wherein Heaps' law can be distinguished between systems that have the same $D(N)$, by considering how pairs of unique functional elements can combine to form new functions.

The challenge with all simple statistical models is in what we can learn about a system \cite{stumpf2012critical}. Recent work has revived the question of the utility of parsimonious models in an era of highly successful non-parsimonious scientific modeling \cite{dubova2025ockham}. 
We contend that our mathematical modeling is useful in the following three ways: (i) Our model allows one to understand that changing the rate at which new functions are created in an organization, or how existing functions are expanded, can have long-term impacts on the structure of the entire system. It may then be possible to design protocols which can tune a system's composition simply by turning two knobs (i.e., $\gamma$ and $\theta$). (ii) Our model allows one to make predictions that have not yet been observed in data, which may be of practical relevance. Examples of this are the self-similarity observed in Fig.~\ref{fig:rank-freq-univ} and the scaling of $n_k$ observed in Figs.~\ref{fig:city-nk} and \ref{fig:si-nk}. (iii) Our model unites previously distinct empirical phenomena, notably explaining the self-similarity of the rank-frequency distributions across cities through their logarithmically increasing diversity. Therefore, we not only have a holistic theory that explains the shapes and forms of abundance distributions and rank-frequency distributions, but we can use our theory to predicts changes in the structure of an organization given policy changes (i.e., changing $\theta$ or $\gamma$). Our model provides new universalities, targets for measurement, and testable predictions. For example, in a growing city, do the rates of function growth inferred from the final distribution match the actual time series for the growth of those functions? This represents a frontier for distributional perspectives in biology and the social sciences, for which our' model provides a new lens.


Considering models from which diversity is an emergent, rather than encoded aspect of a mathematical model, could help to further elaborate on the benefits and drawbacks of diversity in complex systems, and to understand the conditions under which given scalings of diversity occur from a more teleological perspective \cite{jain1998autocatalytic,jain2001model,soyer2010evolution,tatka2025speciated}. Additionally, data analyses in the present study could be extended to further systems, such as biological and computational ecosystems \cite{schueller2022evolving}, and engineered systems such as Airfix models and Lego \cite{bartneck2018lego}. For example, our approach could explain the dynamics of evolution of programming languages, such as Rust, in terms of changes in the number of developers, users, and downloads of packages. Our contribution provides methodological and conceptual steps in these directions.



\section*{Acknowledgements}

\noindent The authors would like to acknowledge the support of the National Science Foundation Grant Award Number EF--2133863. H.Y.~acknowledges the NRF Global Humanities and Social Sciences Convergence Research Program (2024S1A5C3A02042671) and the support from the Institute of Management Research at Seoul National University. J.H.~acknowledges the support of a Lou Schuyler grant from the Santa Fe Institute.

\section*{Author Contributions}
\noindent Conceptualization: J.H., S.R., V.C.Y., G.B.W., C.K., and H.Y. Mathematical analysis: J.H., S.R., and P.L.K. Funding acquisition: S.R., V.C.Y., G.B.W., C.K., and H.Y. Simulations and computational analyses: J.H. Data acquisition: J.I.A., V.C.Y., and H.Y. Supervision: S.R., G.B.W., C.K., and H.Y. Figures: J.H. Writing--original draft: J.H. Writing--review and editing: all authors.
\bibliographystyle{naturemag.bst}
\bibliography{main}

\pagebreak
\clearpage
\widetext
\begin{center}
\textbf{\large Supplementary Information}
\end{center}
\setcounter{equation}{0}
\setcounter{figure}{0}
\setcounter{section}{0}
\setcounter{table}{0}
\setcounter{footnote}{0}
\setcounter{page}{1}
\makeatletter
\renewcommand{\thesection}{S\arabic{section}}
\renewcommand{\theequation}{S\arabic{equation}}
\renewcommand{\thefigure}{S\arabic{figure}}

\section{Special case solutions of the model}\label{sec:exact}

\subsection{Solution for $\gamma=\theta=1$}\label{sec:exact1}
\noindent In this section, we exactly solve Eq.~\eqref{eq:genME} wherein $\theta=\gamma = 1$. Now, we define $p=1/(N+1)$ and $q_k = k/(N+1)$ for system size $N$. The approach in this section introduces a random variable $N_k(N)$, which is the number of functions of size $k$ for a given stochastic trajectory at system size $N$. Connecting to the main text, $\langle N_k(N)\rangle = n_k(N)$. We treat system size $N$ as a discrete variable here, rather than as a continuous variable as in the main text.

\begin{figure}[h!]
\includegraphics[width=0.5\textwidth]{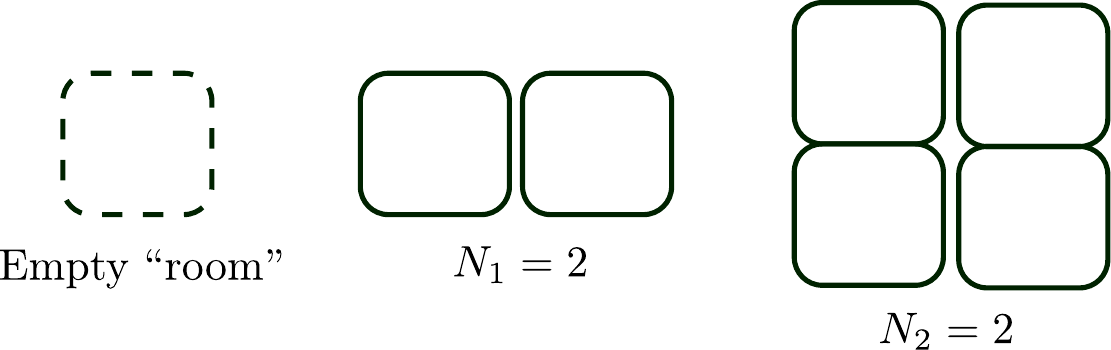}
    \caption{\textbf{Schematic showing an organization with an empty ``room'' that can be added to, two functions of size 1, and two functions of size 2.}}
    \label{fig:exact_11}
\end{figure}

Adding a new function can be thought of as adding to an empty room with probability $1/(N+1)$ (see Fig.~\ref{fig:exact_11}). Existing functions are added to with probability $kN_k/(N+1)$. Increasing the system size, $N\to N+1$, leads to the stochastic equation 
\begin{equation}
\label{N1-rec}
N_1\to 
\begin{cases}
N_1+1, & \text{prob.} \quad \frac{1}{N+1},\\
N_1-1, & \text{prob.} \quad \frac{N_1}{N+1},\\
N_1,   & \text{prob.} \quad 1-\frac{N_1+1}{N+1},
\end{cases}
\end{equation}
for the random variable $N_1$. The initial condition is deterministic, $N_1(1)=1$, while for $N\geq 2$, the quantity $N_1(N)$ is stochastic:
\begin{equation}
\label{N1(2)}
N_1(2) = 
\begin{cases}
2, & \text{prob.} \quad \frac{1}{2},\\
0, & \text{prob.} \quad \frac{1}{2},
\end{cases}
\end{equation}
and so on. The stochastic equation \eqref{N1-rec}, for the random variable $N_1$, is equivalent to the deterministic equation
\begin{align}
\label{PN1N}
P(N_1,N+1) =& \frac{P(N_1-1,N)+(N_1+1)P(N_1+1,N)}{N+1} + \left[1-\frac{N_1+1}{N+1}\right]P(N_1,N)
\end{align}
for the probability distribution $P(N_1,N)$. We do not know the general solution of the difference equation \eqref{PN1N}. In the most interesting case, the $N\to\infty$ limit, the probability distribution becomes `stationary', i.e., asymptotically independent on $N$ playing a role of time.  The stationary distribution $P(N_1)=P(N_1,\infty)$ satisfies 
\begin{equation}
\label{PN1-inf}
P(N_1-1) + (N_1+1)P(N_1+1) = (N_1+1)P(N_1)
\end{equation}
which we deduce from Eq.~\eqref{PN1N}. The solution of \eqref{PN1-inf} is the Poisson distribution
\begin{equation}
\label{PN1-sol}
P(N_1) = \frac{e^{-1}}{N_1!}.
\end{equation}
Using \eqref{PN1-sol} we compute the average
\begin{equation}
\label{N1-av}
\langle N_1(\infty)\rangle=\sum_{N_1\geq 0}N_1\,\frac{e^{-1}}{N_1!}=1.
\end{equation}
A remarkable property of the Poisson distribution is that all cumulants are equal to the average. Therefore the cumulants are $\langle N_1^p(\infty)\rangle_c=1$ for all integer $p\geq 1$. 

An amusing feature of the random variable $N_1(N)$ is that its average is perfectly stationary from the very beginning, i.e., for all $N\geq 1$. This is valid for $N=1$ since $N_1(1)=1$, and it can be checked for $N=2$ by averaging \eqref{N1(2)}. Generally, the average satisfies the recurrence
\begin{equation}
\label{N1N-rec}
\langle N_1(N+1)\rangle = \frac{N}{N+1}\,\langle N_1(N)\rangle+\frac{1}{N+1}.
\end{equation}
One can verify that Eq.~\eqref{N1N-rec} admits the stationary solution $\langle N_1(N)\rangle=1$ for all $N\geq 1$. The perfect stationarity of the average is a much stronger property than the asymptotic stationarity of the distribution $P(N_1, N)$. For any finite $N$, all values of $P(N_1, N)$ are rational rather than transcendental values of the limiting distribution \eqref{PN1-sol}. A more elementary manifestation of an asymptotic rather than perfect stationarity of $P(N_1, N)$ is the vanishing property: $P(N_1, N)=0$ for $N_1>N$, while the limiting distribution \eqref{PN1-sol} is positive for all $N_1\geq 0$.

Similarly to the average, one computes the variance and finds that it also quickly becomes perfectly stationary: $\langle N_1^2(N)\rangle_c=1$ for $N\geq 2$. This remarkable perfect stationarity holds for all cumulants and merely settles on later for higher cumulants:
\begin{equation}
\label{N1-cum}
\langle N_1^p(N)\rangle_c = 1  \qquad\text{for}\quad N\geq p.
\end{equation}

Let us now look at the next random variables $N_2$. Increasing the system size, $N\to N+1$, leads to the stochastic equation
\begin{equation}
\label{N2-rec}
N_2\to 
\begin{cases}
N_2+1, & \mathrm{prob.} \quad \frac{N_1}{N+1},\\
N_2-1, & \mathrm{prob.} \quad \frac{2N_2}{N+1},\\
N_2,    & \mathrm{prob.} \quad 1-\frac{N_1+2N_2}{N+1}. 
\end{cases}
\end{equation}
The evolution of $N_1$ is self-consistent, see \eqref{N1-rec} or  \eqref{PN1N}. The evolution of $N_2$ depends on $N_1$. Therefore instead of \eqref{N2-rec} it is better to consider the evolution of the pair $(N_1,N_2)$. The increase of the system size, $N\to N+1$, leads to the stochastic equation 
\begin{equation*}
\label{N12-rec}
(N_1,N_2)\to 
\begin{cases}
(N_1+1,N_2),    & \mathrm{prob.} \quad \frac{1}{N+1},\\
(N_1-1,N_2+1), & \mathrm{prob.} \quad \frac{N_1}{N+1},\\
(N_1,N_2-1),    & \mathrm{prob.} \quad \frac{2N_2}{N+1}, \\
(N_1,N_2),       & \mathrm{prob.} \quad 1-\frac{1+N_1+2N_2}{N+1}. 
\end{cases}
\end{equation*}
The probability distribution $P(N_1,N_2, N)$ also becomes stationary in the $N\to\infty$ limit. The stationary distribution $P(N_1,N_2)=P(N_1,N_2,\infty)$ satisfies 
\begin{align}
\label{PN12-inf}
(1+N_1+2N_2)P(N_1,N_2) = 2(N_2+1)P(N_1,N_2+1) 
+ (N_1+1)P(N_1+1,N_2-1)
+ P(N_1-1,N_2)
\end{align}
deduced from an equation for $P(N_1,N_2,N)$ similarly to the derivation of Eq.~\eqref{PN1-inf} from Eq.~\eqref{PN1N}. The complicated difference equation \eqref{PN12-inf} has a neat factorizable solution 
\begin{equation}
\label{PN12-sol}
P(N_1,N_2) = \frac{e^{-3/2}}{N_1! N_2! \,2^{N_2}}
\end{equation}
which is a product of the two Poisson distributions. 

Thus, $N_1(\infty)$ and $N_2(\infty)$ are uncorrelated. The random quantities $N_1(N)$ and $N_2(N)$ are correlated. Let us look at the average of the product, $\langle N_1(N)N_2(N)\rangle$. The initial condition is $N_1(1)=1, N_2(1)=0$. Thus $\langle N_1(1)N_2(1)\rangle=0$ and straightforward calculations give $\langle N_1(2)N_2(2)\rangle=0$ and $\langle N_1(3)N_2(3)\rangle=\frac{1}{2}$. The first non-vanishing value is the ultimate stationary value: $\langle N_1(N)N_2(N)\rangle=\frac{1}{2}$ for all $N\geq 3$. This assertion can be established by solving an equation for $\langle N_1(N)N_2(N)\rangle$. The perfect stationarity of the average product is a specific case of the perfect stationarity of the moments 
\begin{equation}
\label{perfect}
\langle N_1^p(N)N^q_2(N)\rangle = \langle N_1^p(\infty)\rangle \langle N^q_2(\infty)\rangle
\end{equation}
which is expected to hold for all integer $p, q\geq 0$ when $N$ is sufficiently large: $N\geq n(p,q)$. We do not have a general proof of the perfect stationarity \eqref{perfect}. We also do not know an analytical formula for the threshold size $n(p,q)$ when the perfect stationarity settles on; conjecturally, $n(p,q)=p+q+1$.

Continuing, we consider $N_1(N),\ldots, N_k(N)$ and notice that the probability distribution $P(N_1,\ldots,N_k,N)$ becomes stationary in the $N\to\infty$ limit. The stationary distribution $P(N_1,\ldots,N_k)=P(N_1,\ldots, N_k,\infty)$ is a product of $k$ Poisson distributions:
\begin{equation}
\label{PN1k-sol}
P(N_1,\ldots,N_k) = \left\{e^{H_k}\prod_{j=1}^k j^{N_j} N_j!\right\}^{-1}
\end{equation}
where $H_k$ is the harmonic number. The perfect stationarity of the  moments 
\begin{equation}
\left \langle \prod_{j=1}^ k N_j^{p_j}(N)\right\rangle = \prod_{j=1}^ k \left \langle  N_j^{p_j}(\infty)\right\rangle 
\end{equation}
is expected to hold for all integer $p_1, \ldots,p_k \geq 0$ when $N$ is sufficiently large: $N\geq n(p_1,\ldots, p_k)$. 

The diversity is a random variable 
\begin{equation}
\mathcal{D}(N)=\sum_{k=1}^N N_k(N),
\end{equation}
where $\langle \mathcal{D}(N)\rangle = D(N)$ in the main text. Increasing the system size, $N\to N+1$, leads to the stochastic equation 
\begin{equation}
\label{D-rec}
\mathcal{D}\to 
\begin{cases}
\mathcal{D}+1, & \mathrm{prob.} \quad \frac{1}{N+1},\\
\mathcal{D},     & \mathrm{prob.} \quad \frac{N}{N+1},
\end{cases}
\end{equation}
from which we deduce the governing equation for the probability distribution $\Pi(D,N)$ of the diversity
\begin{eqnarray}
\label{PDN}
\Pi(\mathcal{D},N+1) = \frac{\Pi(\mathcal{D}-1,N)+N\Pi(\mathcal{D},N)}{N+1}.
\end{eqnarray}
The solution of this recurrence is expressible via the Stirling numbers \cite{Stirling,Knuth} of the first kind:
\begin{equation}
\label{R2:Stirling}
\Pi(\mathcal{D},N)=\frac{1}{N!}\,{N\brack \mathcal{D}}.
\end{equation}
The average diversity 
\begin{equation}
\langle \mathcal{D}(N)\rangle = \sum_{\mathcal{D}=1}^N \mathcal{D} \Pi(\mathcal{D},N)= H_N
\end{equation}
can be derived directly from \eqref{PDN} or from the generating function
\begin{equation}
\sum_{\mathcal{D}=1}^N x^\mathcal{D} \Pi(\mathcal{D},N) = \frac{1}{N!}\,\prod_{j=0}^{N-1}(x+j)
\end{equation}
which essentially defines  the Stirling numbers of the first kind \cite{Knuth}. 
We also mention a few extremal probabilities
\begin{equation*}
\begin{split}
& \Pi(1,N)=N^{-1}\,, \qquad \qquad\quad  \Pi(2,N)=N^{-1}H_{N-1} \\
&  \Pi(N-1,N)=\frac{1}{N!}\binom{N}{2}\,,  \quad \Pi(N,N)=\frac{1}{N!}
\end{split}
\end{equation*}
following from identities for extremal values of the Stirling numbers of the first kind \cite{Knuth}.

\subsection{Solution for $\gamma=\theta=0$}\label{sec:exact2}
\noindent In this model, the diversity evolves according to 
\begin{equation}
\label{D0}
\mathcal{D}\to 
\begin{cases}
\mathcal{D}+1, & \text{prob.} \quad \frac{1}{\mathcal{D}+1},\\
\mathcal{D},     & \text{prob.} \quad \frac{\mathcal{D}}{\mathcal{D}+1}, 
\end{cases}
\end{equation}
when the system size increases, $N\to N+1$. This stochastic equation is much more challenging than the corresponding equation in the previous model since the probabilities in Eq.~\eqref{D0} are random while they are deterministic in Eq.~\eqref{D-rec}. Averaging Eq.~\eqref{D-rec} we deduce
\begin{equation}
\label{DN-rec}
\langle \mathcal{D}(N+1)\rangle = \langle \mathcal{D}(N)\rangle + \delta
\end{equation}
where we shortly write $\delta=\langle [\mathcal{D}(N)+1]^{-1}\rangle$. Equation \eqref{DN-rec} is exact, but analytically intractable since $\delta$ cannot be expressed via $\langle \mathcal{D}(N)\rangle$, so we don't have a closed recurrence for $\langle \mathcal{D}(N)\rangle$. 

However, we are interesting in the large $N$ behavior and we now show how to extract the leading asymptotic behavior of $\langle \mathcal{D}(N)\rangle$. The derivation relies on the key property of the random variable $\mathcal{D}(N)$, viz., its asymptotic self-averaging giving $\delta\simeq \frac{1}{\langle \mathcal{D}(N)\rangle}$ for $N\gg 1$. We replace the difference $\langle \mathcal{D}(N+1)\rangle - \langle \mathcal{D}(N)\rangle$ by the derivative $\frac{d \langle \mathcal{D}(N)\rangle}{dN}$ for $N\gg 1$, so \eqref{DN-rec} becomes $\frac{d \langle \mathcal{D}(N)\rangle}{dN}\simeq \frac{1}{\langle \mathcal{D}(N)\rangle}$ from which
\begin{equation}
\label{DN-sol}
\langle \mathcal{D}(N)\rangle \simeq \sqrt{2N}
\end{equation}
for $N\gg 1$. 

To justify the asymptotic self-averaging of the random variable $\mathcal{D}(N)$ in a constructive way we consider the variance $\sigma(N)=\langle \mathcal{D}^2(N)\rangle_c$. Similarly to Eq.~\eqref{DN-rec} one derives an equation
\begin{equation}
\label{sigma-N-rec}
\sigma(N+1) = \sigma(N) + 2\left[1-\langle \mathcal{D}(N)\rangle \delta\right]- \delta - \delta^2
\end{equation}
The asymptotic behavior of the average \eqref{DN-sol} suggests that the deviation from the average scales as $N^\frac{1}{4}$. Therefore we write $D(N)$ as the sum of the deterministic and stochastic component
\begin{equation}
\label{DN-stoch}
\mathcal{D}(N) = \langle \mathcal{D}(N)\rangle + N^\frac{1}{4}\xi
\end{equation}
The random variable $\xi$ has zero mean, $\langle \xi \rangle = 0$, and variance $\langle \xi^2 \rangle = \sigma_0$ implying that $\sigma(N) =  \sigma_0 \sqrt{N}$. Using \eqref{DN-stoch} we compute
\begin{equation}
\label{delta}
\delta = \left\langle \frac{1}{\langle \mathcal{D}(N)\rangle + N^\frac{1}{4}\xi+1}\right\rangle \simeq  \frac{1}{\langle \mathcal{D}(N)\rangle +1} + \frac{\sigma_0 \sqrt{N}}{(\langle \mathcal{D}(N)\rangle +1)^3}
\end{equation}
Substituting \eqref{delta} into \eqref{sigma-N-rec}, employing a continuum approximation, using \eqref{DN-sol}, and keeping only asymptotically dominant terms we simplify \eqref{sigma-N-rec} to 
\begin{equation}
\label{sigma-N-eq}
\frac{\mathrm{d}}{\mathrm{d}N}\left(\sigma_0 \sqrt{N}\right)=\frac{1}{\sqrt{2N}}-\frac{\sigma_0}{\sqrt{N}}
\end{equation}
which we solve to fix $\sigma_0=\sqrt{2}/3$. Thus 
\begin{equation}
\label{sigma-N-sol}
\sigma(N) \simeq \frac{1}{3}\sqrt{2N}.
\end{equation}

For $N_k$, one can show that its mean $\langle N_k \rangle$ has the following evolution equation,
\begin{align}
    \langle N_k(N+1)\rangle = \langle N_k(N)\rangle + \frac{\delta_{k,1}+\langle N_{k-1}(N)\rangle-\langle N_{k}(N)\rangle }{\langle \mathcal{D}(N)\rangle +1},\quad k\geq 1,
\end{align}
in which $\langle N_{0}(N)\rangle = 0$, and where we have only kept dominant terms from Eq.~\eqref{delta}. Assuming $N$ is continuous, this expression simplifies to the following ordinary differential equation
\begin{align}\label{eq:mid_00}
    \frac{\mathrm{d} \langle N_k(N)\rangle}{\mathrm{d} N}= \frac{\delta_{k,1}+\langle N_{k-1}(N)\rangle-\langle N_{k}(N)\rangle }{\langle \mathcal{D}(N)\rangle +1}.
\end{align}
In the regime of large $N$ this ordinary differential equation admits a useful solution. We now use $\langle \mathcal{D}(N)\rangle+1\sim N^{1/2}$, and employ the independent variable transform $x = N^{1/2}$, wherein $\mathrm{d}x = \mathrm{d}N N^{-1/2}/2$, transforming Eq.~\eqref{eq:mid_00} to
\begin{align}
    \frac{\mathrm{d} \langle N_k(x)\rangle}{\mathrm{d} x} \approx \delta_{k,1}+\langle N_{k-1}(x)\rangle-\langle N_{k}(x)\rangle.
\end{align}
These equations are readily solved one-by-one, to give
\begin{align}\nonumber
     \langle N_1(x)\rangle &= 1-e^{-x},\\
     \langle N_2(x)\rangle &= 1-(1+x)e^{-x},
\end{align}
which in the general case gives,
\begin{align}
    \langle N_k(N)\rangle &= 1-e^{-\sqrt{N}}\sum_{j=0}^{k-1}\frac{N^{j/2}}{j!}.
\end{align}
Therefore, in the limit of large $N$, $\langle N_k(N)\rangle =1$. Because $\langle \mathcal{D}(N)\rangle\sim \sqrt{N}$ and $\sum_{k=1}^{k_{\mathrm{max}}(N)} \langle N_k(N)\rangle = \langle\mathcal{D}(N)\rangle$, this also implies that the size of the largest function scales as $k_{\mathrm{max}}(N)\sim \sqrt{N}$. In Section \ref{sec:sublinpa}, we also show that the scaling of the largest function is $\sqrt{N}$.


\section{Derivation of $\sum_k k^x n_k(N)\sim N^{\delta(x)}$ for large $N$}\label{sec:ansatz}

\noindent In this section we prove that $\sum_k k^x n_k(N)\sim N^{\delta(x)}$ for large $N$. First, we multiply Eq.~\eqref{eq:genME} by $k^x$ and then sum over $k$ to give,
\begin{align}
    \partial_N \sum_k k^xn_k = p(\theta,N) + \frac{\sum_k \left[k^x(k-1)^\gamma n_{k-1} - k^{x+\gamma}n_k\right]}{\sum_k k^\gamma n_k}.
\end{align}
In the limit of $k\gg 1$ one can write the numerator of the second term as $k^{x+\gamma}(n_{k-1}-n_k)$. We can then perform the sums in $n_{k-1}$ and $n_k$ separately. For the sum including $n_{k-1}$ we get,
\begin{align}
    \sum_k k^{x+\gamma}n_{k-1} = \sum_{k} (k+1)^{x+\gamma}n_k \approx \sum_k k^{x+\gamma}n_k\left[ 1 +\frac{x+\gamma}{k} \right],
\end{align}
in which the final approximation is valid in the limit $k\gg1$. Defining $f(x,N) = \sum_k k^x n_k(N)$ we find
\begin{align}\label{eq:f_eq}
    \partial_N f(x,N) \approx \frac{f(x+\gamma-1,N)}{f(\gamma,N)},
\end{align}
using the fact that $p(\theta,N\gg 1) \ll 1$. A simple non-trivial function that satisfies Eq.~\eqref{eq:f_eq} is $f(x,N) = N^{\delta(x)}$. Substituting this functional form into Eq.~\eqref{eq:f_eq}, we obtain an equation determining $\delta(x)$, given by $$\delta(x)-1 = \delta(x+\gamma-1)-\delta(\gamma).$$ There will be a solution to this equation if $\delta(x)$ is a linear function of $x$. Taking $\delta(x) = a x +b$, one can show that $\delta(x) = a x +(1-a)$ in which $a$ is a constant to be determined. This is consistent with the relation $f(1,N)=N$ or $\sum_k k n_k = N$. To determine $a$, we can use the two equivalent definitions of $D(N)$ given by
$$D(N) = \sum_k n_k \sim N^{\delta(0)},$$
and noting that $p = p_0/(\sum_k k^\theta n_k)$ we also find
$$D(N) = \int^N p(\theta,N')\mathrm{d}N' \sim N^{1-\delta(\theta)},$$
using $p(\theta,N) \sim f(\theta,N)^{-1}$. This equivalence results in the relation $\delta(0) = 1-\delta(\theta)$ which returns $a = 1/(2-\theta)$, giving the general asymptotic result
\begin{align}
    \sum_k k^x n_k(N) \sim N^{\frac{x-1}{2-\theta}+1}.
\end{align}
This completes the derivation of Eq.~\eqref{eq:sum-scaling} in the main text.


\section{Derivation of the leading terms in $n_k(N)$}\label{sec:leading_deriv}
\noindent Starting from Eq.~\eqref{eq:genME}, we can write out the terms of the master equation as follows
\begin{align}\nonumber
    \frac{\partial n_1(N)}{\partial N} &= p - (1-p) q_1 n_1(N),\\\nonumber
    \frac{\partial n_k(N)}{\partial N} &= (1-p)(q_{k-1} n_{k-1}(N) -q_k n_k(N)), \quad k>1.
\end{align}
In the regime of sub-linear preferential attachment $\partial_N n_k(N)\ll q_k n_k(N)$ (by definition), and following dominant balance \cite[p.~83]{bender2013advanced}, one can then solve these equations iteratively to give the leading order terms of $n_k(N)$ in $N$:
\begin{align}\nonumber
    n_1(N) &\sim \frac{p}{(1-p)q_1} = \frac{p}{q_1}+\mathcal{O}(p^2),\\\nonumber
    n_k(N) &\sim \frac{p}{(1-p)q_k} = \frac{p}{q_k}+\mathcal{O}(p^2), \quad k>1.
\end{align}
In the asymptotic limit $p^2\ll p$, and therefore higher-order terms of $p$ can be neglected. Using the scaling of $p\sim N^{-\nu}$ and $q_k \sim k^\gamma/N^\mu$ from the main text, this leads to $n_k(N) \sim k^{-\gamma} N^{\mu-\nu}$. This form of $n_k$ can be easily shown to satisfy $\partial_N n_k(N)\ll q_k n_k(N)$ for $\gamma<1$, which validates the usage of the dominant balance approximation.

In the regime of linear preferential attachment, by definition, $\partial_N n_k(N) \sim k n_k(N)/N$ for $k>1$, i.e., the rate of growth of any function is proportional to its size. Therefore, the dominant balance approximation $\partial_N n_k(N)\ll n_k(N)$ no longer holds for linear preferential attachment. However, it is still true that $\partial_N n_1(N) \ll n_1(N)$---since the mechanism of adding new functions remains the same---which leads to $n_1(N)\sim p N\sim N^{1-\nu}$. Using this functional form as an ansatz for general $n_k$, i.e., $n_k(N)\sim a_k N^{1-\nu}$, we can then substitute this ansatz into Eq.~\ref{eq:genME} to obtain
\begin{align}
    (1-\nu)a_k \sim ((k-1)a_{k-1}-k a_k) + \delta_{k,1}.
\end{align}
Solving this two-term recurrence relation for $a_k$ then gives us $n_k(N)$ for linear preferential attachment,
\begin{align}
    n_k(N) \sim k^{\nu-2}N^{1-\nu}.
\end{align}
Curiously, each $n_k(N)$ is now directly proportional to the diversity of the system, since $D(N)\sim N^{1-\nu}$. This has implications for the self-similarity observed in the case of linear preferential attachment explored in Section \ref{sec:linpa}.

\clearpage
\section{Verification of functional form of $n_k(N)$}
\begin{figure}[h!]
\includegraphics[width=0.75\textwidth]{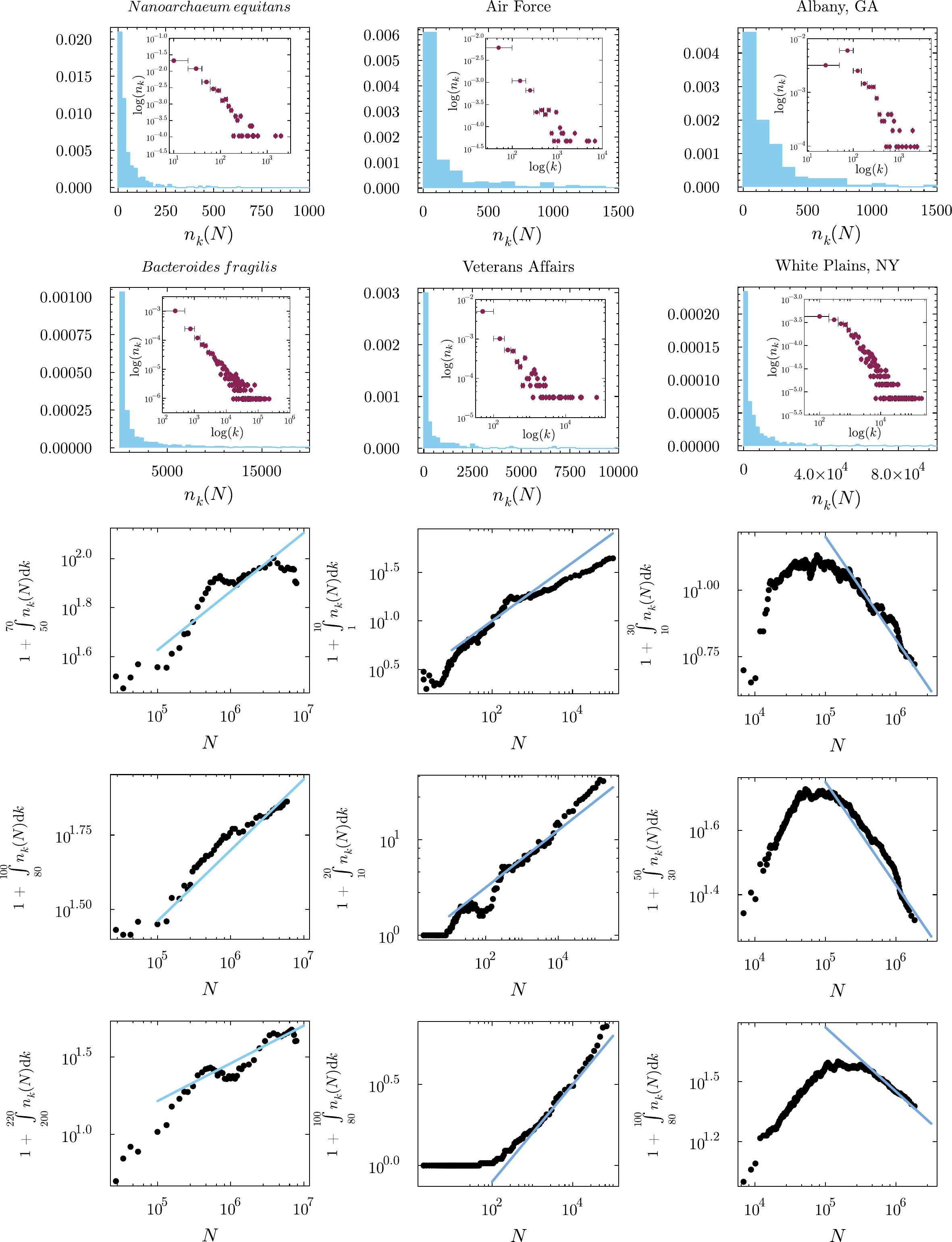}
    \caption{\textbf{Exhibition of scaling in the data for comparison to the form of $n_k(N)\sim k^{-\gamma}N^{(\gamma-\theta)/(2-\theta)}$.} The first column show data from the prokaryotes, the second for the federal agencies, and the third from the cities. The first two rows show examples from the data of the power-law dependence of $n_k(N)$, for fixed $N$, across all three datasets. The last three rows show the power-law dependence of $n_k(N)$, against $N$, for a fixed set of values $k$ (evaluated as an integral due to sparsity of data). To smooth the data, each data point is a rolling average over 10 data points for the prokaryotes, 20 data points for the federal agencies, and 30 data points for the cities. In each case, the blue solid line shows the same power-law dependence for each system, which shows the independence of the scaling of $n_k(N)$ in $N$ for different values of $k$ ($N^{0.21}$ for the prokaryotes, $N^{0.40}$ for the federal agencies, and $N^{-0.32}$ for the cities). Note that functions of large size $k$ require a given size $N$ before the predicted scaling with respect to $N$ begins to hold, which is also seen in the simulations in Fig.~\ref{fig:conf-analytics}(d), which is most apparent in the case  of cities due to the decreasing power law dependence of $n_k(N)$ on $N$.}
    \label{fig:si-nk}
\end{figure}

\clearpage
\section{Effect of a minimal set of functions}\label{sec:minimal-set}
\noindent In the main text we assume that organizations evolve from the initial condition $n_k(1)=\delta_{k,1}$. Although the functional form of the scaling laws and the predicted self-similarities are independent of the initial condition in the regime of large $N$, it can be useful to assess: (1) How an initial condition $n_k(1)\neq\delta_{k,1}$ manifests itself in the scaling laws, rank-frequency distributions, and self-similarity conditions. (2) How the scaling and functional form of the diversity scaling is modified starting from $n_k(1)\neq\delta_{k,1}$.

\begin{figure}[h!]
    \includegraphics[width=0.73\textwidth]{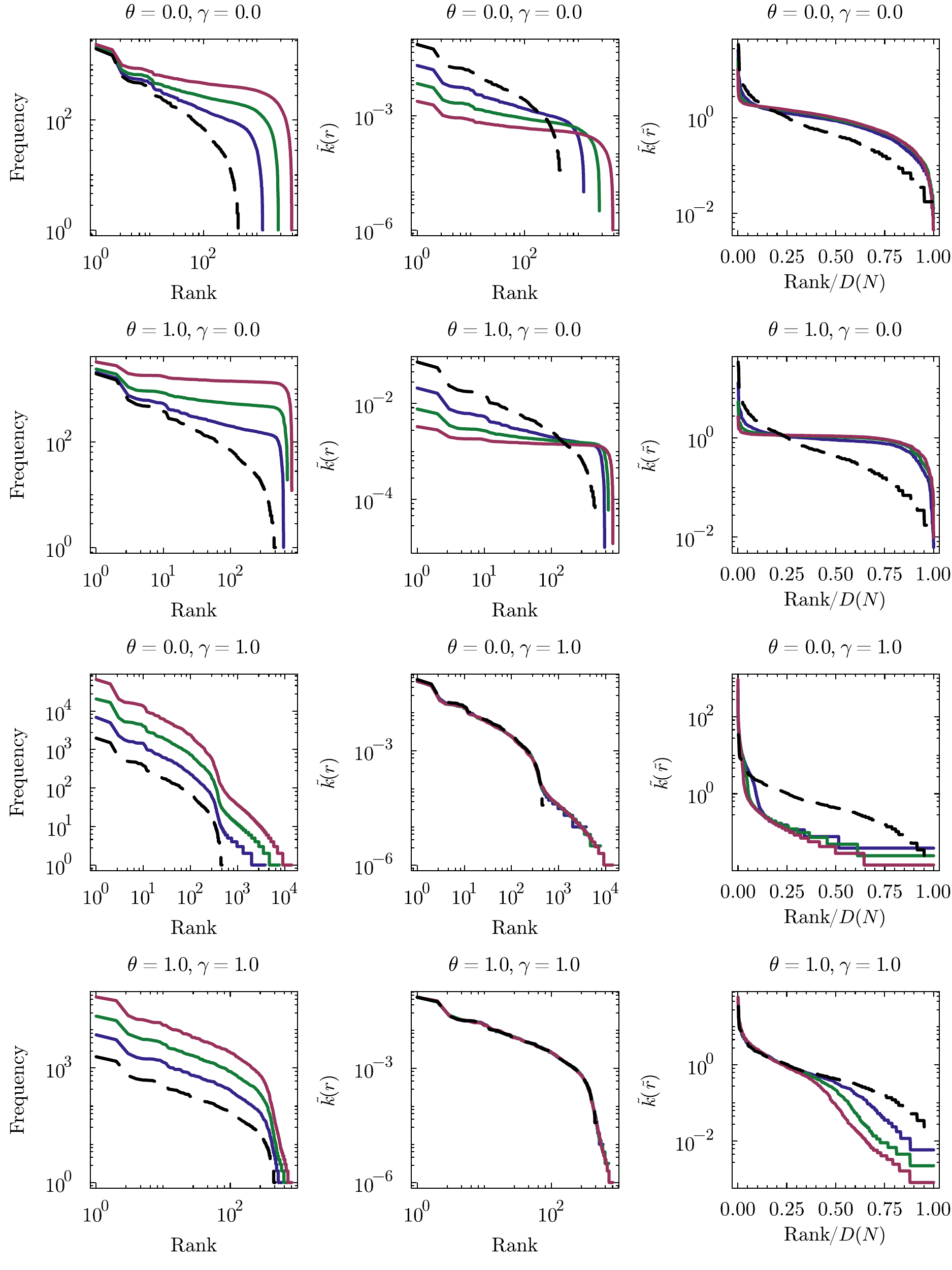}
    \caption{\textbf{Simulations showing the effect of a minimal set of functions in which $n_k(1)\neq \delta_{k,1}$.} Each row shows a different case of the data. The first column shows the rank-frequency distribution unnormalized, the second column shows the normalized rank-frequency distribution $\tilde{k}(r)$, and the third column shows the rank-normalized rank-frequency distribution $\tilde{k}(\tilde{r})$. The black dashed line shows the initial distribution of functions, which is the data from the smallest prokaryotic cell of size $N=26,925$. The blue, green and red lines show the distributions for $N=10^5,10^{5.5}$ and $10^6$ respectively. The parameters $\theta$ and $\gamma$ are provided in the title of each plot.}
    \label{fig:non-naive}
\end{figure}

To answer point (1), in Fig.~\ref{fig:non-naive} we simulated four different parameter sets starting from an initial set of functions corresponding to the rank-frequency distribution of proteins for the smallest prokaryotic cell. These simulations show that: (i) The predicted self-similarity of $\tilde{k}(\tilde{r})$, for $\gamma<1$ and $\theta<1$, is still exhibited (top row). (ii) The self-similarity of $\tilde{k}(\tilde{r})$ and $\Tilde{k}(r)$ for $\gamma<1$ and $\theta=1$ is still exhibited (second row), with the self-similarity in $\Tilde{k}(r)$ converging more slowly than the self-similarity in $\Tilde{k}(\tilde r)$. (iii) There is self-similarity in $\Tilde{k}(r)$, and lack of it in $\tilde{k}(\tilde{r})$, for $\gamma=1$ (bottom two rows).

For point (2), consider starting with a system of size $N_0$ given an initial condition $n_k(1)\neq\delta_{k,1}$. This scaling of diversity will then take the form
\begin{align}
    D(N) = D(N_0)+ \int^N_{N_0} p(\theta,N')\mathrm{d}N' \approx D(N_0) + \kappa\left[N^{\frac{1-\theta}{2-\theta}}-N_0^{\frac{1-\theta}{2-\theta}} \right],
\end{align}
since the scaling for $p$ will still hold in the asymptotic regime and $\theta<1$, and $\kappa$ is a constant coming from $p = p_0/\sum_k k^\theta n_k(N)\sim  \kappa\,N^{-1/(2-\theta)}$. Similar results can be derived for $\theta=1$. If we know the diversity and system size at one other system size, say $D(N^\star)$, then we can use this to eliminate $\kappa$ to give the result
\begin{align}
    D(N) \approx D(N_0) + \frac{\left[N^{\frac{1-\theta}{2-\theta}}-N_0^{\frac{1-\theta}{2-\theta}} \right]\left[D(N^{\star})-D(N_0) \right]}{{N^{\star^{\frac{1-\theta}{2-\theta}}}-N_0^{\frac{1-\theta}{2-\theta}}}}.
\end{align}
This gives a shifted form of the diversity scaling, that has the same scaling behavior as that seen for the diversity in the main text.

\end{document}